\documentclass[a4paper,11pt]{article}
\pdfoutput=1
\usepackage{jcappub}

\def\prd{Phys. Rev. D}
\def\mnras{MNRAS}
\def\apj{ApJ}
\def\apjl{ApJL}

\def\aap{A\&A}

\def\aapr{A\&A Rev.}

\def\jcap{JCAP}

\usepackage{amssymb}		 
\usepackage{amsmath, amsthm, amssymb,amsfonts}
\usepackage[english]{babel}
\usepackage{graphicx}
\usepackage{epstopdf} 
\usepackage{booktabs}
\usepackage{comment}

\newcommand\uv{{\bf u}}
\newcommand\bbv{{\bf b}}

\newcommand\xv{{\bf x}}

\newcommand\Bv{{\bf B}}

\newcommand\bnabla{\boldsymbol{\nabla}}

\begin{document}

\begin{flushright}
TUM-HEP 1089/17
\end{flushright}

\title{A signature of anisotropic cosmic-ray transport in the gamma-ray sky}

\author[a]{Silvio Sergio Cerri}
\author[b]{Daniele Gaggero}
\author[c]{Andrea Vittino}
\author[d]{Carmelo Evoli}
\author[a,e]{Dario Grasso}

\affiliation[a]{Dipartimento di Fisica ``E. Fermi'', Universit\`{a} di Pisa, Largo B. Pontecorvo 3, I-56127 Pisa, Italy}
\affiliation[b]{GRAPPA, University of Amsterdam, Science Park 904, 1098 XH Amsterdam, Netherlands}
\affiliation[c]{Physik-Department T30d, Technische Universit{\"a}t M{\"u}nchen, James Franck-Str. 1, D-85748, Garching, Germany}
\affiliation[d]{Gran Sasso Science Institute, Viale Francesco Crispi 7, 67100  L'Aquila, Italy}
\affiliation[e]{INFN Pisa, Largo B. Pontecorvo 3, I-56127 Pisa, Italy}

\emailAdd{silvio.cerri@df.unipi.it}
\emailAdd{d.gaggero@uva.nl}
\emailAdd{andrea.vittino@tum.de}
\emailAdd{carmelo.evoli@gssi.it}
\emailAdd{dario.grasso@pi.infn.it}

\abstract{
A crucial process in Galactic cosmic-ray (CR) transport is the spatial diffusion due to the interaction with the interstellar turbulent magnetic field. 
Usually, CR diffusion is assumed to be uniform and isotropic all across the Galaxy. However, this picture is clearly inaccurate: Several data-driven and theoretical arguments,  {as well as} dedicated numerical simulations, show that diffusion  {exhibits} highly anisotropic properties  {with respect to the} direction of  {a background (ordered)} magnetic field  {(i.e., parallel or perpendicular to it)}. 

In this paper we focus on a recently discovered anomaly in the hadronic CR spectrum inferred by the Fermi-LAT gamma-ray data at different positions in the Galaxy, i.e. the progressive hardening of the proton slope at low Galactocentric radii. We propose the idea that this feature can be interpreted as a signature of anisotropic diffusion  {in the complex Galactic magnetic field}: In particular, the harder slope in the inner Galaxy is due, in our scenario, to the parallel diffusive escape along the poloidal component of the large-scale, regular, magnetic field.

We implement this idea in a numerical framework, based on the {\tt DRAGON} code, and perform detailed numerical tests on the accuracy of our setup. We discuss how the effect proposed depends on the relevant free parameters involved.
Based on low-energy extrapolation of the few focused numerical simulations aimed at determining the scalings of the anisotropic diffusion coefficients, we finally present a set of plausible models that reproduce the behavior of the CR proton slopes inferred by gamma-ray data. 
}

\maketitle

\section{Introduction}

The propagation of charged cosmic rays (CRs) in the Galaxy and their complex interactions with the interstellar medium (ISM) is generally described in terms of a transport equation characterized by a spatially uniform and isotropic diffusion coefficient, $D_{ij}=D\,\delta_{ij}$. 
This quantity has a power-law scaling with the rigidity $p/Z$, i.e., $D\propto(p/Z)^\delta$. This phenomenological approach dates back to the pioneering works by Ginzburg and Syrovatskii in the early 1960s \cite{ginzburg1964}, and was considered adequate to reproduce the available experimental data for a long time \cite{strong2007}. 

Nowadays, this picture is severely challenged both by theoretical arguments and new experimental results. 

A plethora of new measurements featuring unprecedented accuracy and several anomalies, both in the charged CR and gamma-ray measured spectra, call for a profound revision of the conventional scenario (see, e.g., \cite{amato2017} for a very recent review). The non-local observables, {\it in primis} gamma rays, play a major role in this context. In particular, two recent model-independent analyses \cite{Acero2016ApJS, Yang:2016jda} showed that Fermi-LAT data point towards  a gradient in the CR proton slope with respect to the Galactocentric radius, $R$. This feature can be interpreted in a natural way in terms of a radially-dependent scaling of the  isotropic diffusion coefficient with rigidity, namely $D\,\propto\,(p/Z)^{\delta(R)}\,$.  

Recently, a consistent model reproducing both local and non-local data, based on this idea, was presented in \cite{Gaggero:2014xla}. 
Besides Fermi-LAT data, a similar model was shown to consistently reproduce gamma-ray measurement at higher energies performed by Milagro \cite{Gaggero:2015xza} and H.E.S.S. \cite{Gaggero:2017jts}, providing a viable solution to well-known discrepancies between those results and conventional models. 

However, the physical picture behind this behavior is still unclear.

On the theoretical side, there are strong reasons to move beyond the simplistic picture of homogeneous and isotropic diffusion as well. 
As a reference theoretical framework for CR transport modeling, one often considers the quasi-linear theory (QLT) of pitch-angle scattering in a random magnetic field  due to Alfv\'en wave packets~\cite{Morrison1957,Jokipii1966,Jokipii1968}. The fundamental assumption behind this scenario is that the power associated to the turbulent fluctuations of the magnetic field, $\delta B$, is much smaller than the one associated to the regular field $B_0$, i.e., $(\delta B/B_0)^2\ll\,1$.
In the ISM, the turbulent spectrum is believed to be injected by supernova explosions at scales of the order $\ell_c \sim 100$ pc, where $\delta B/B_0\sim 1$, or generated  by CR themselves through streaming instability. The dominant contribution to CR diffusion is provided by resonant scattering at the scale of the order of the particle Larmor radius, which, e.g., for rigidities of $\sim1$~GeV, corresponds roughly to an astronomical unit ($10^{13}~{\rm cm}\simeq5\times10^{-6}$~pc).  
Since a turbulent cascade is taking place, at that small scale the turbulent power is extremely suppressed with respect to the one at the scale of the injection.
Therefore the QLT is expected to hold in the environment of our Galaxy, as far as Galactic propagation of particles in the GeV--PeV range is concerned (see, e.g., the detailed discussion in \cite{DeMarco2007a}).
The crucial prediction of such theory is a {\it highly anisotropic} transport regime. 
In fact, the QLT treatment predicts that the ratio between the perpendicular and parallel diffusion coefficients is
\begin{equation}\label{eq:intro4}
\frac{D_\perp}{D_\|}\,\sim\,{\cal F}(k)\,\sim\,\frac{\delta B_k^2}{B_0^2}\,\ll\,1\,.
\end{equation}
where  $D_\|$ and $D_\perp$ are the diffusion coefficients parallel and perpendicular to $\Bv_0$, respectively, and $\mathcal{F}(k)$ is defined as the (normalized) power associated to the turbulent modes with wave number $k \propto 1/p$ resonating with the particles carrying momentum $p$. 

Other processes could actually be at work and complicate this picture, without altering the bottom line, i.e. the need for anisotropic diffusion. In particular, we point out the {\it field line random walk}, which has been widely discussed in the literature and may be able to  enhance the perpendicular diffusion with respect to what is predicted by QLT, boosting the perpendicular diffusion coefficient $D_\perp$ up to $\sim 10\%$ of its parallel counterpart  (see, e.g., \cite{1999ApJ...520..204G}). 

As far as the rigidity dependences of the parallel and perpendicular coefficients are concerned, a useful insight comes from the numerical simulations designed to compute the trajectory of charged test particles in turbulent magnetic fields \cite{2002PhRvD..65b3002C,DeMarco2007a,Snodin2016}. These simulations clearly outline a significantly steeper rigidity dependence for $D_\perp$ with respect to $D_\parallel$. However, these results apply for Larmor radii significantly larger than those relevant for our analysis (i.e. those corresponding to GV-TV rigidities), so a low-energy extrapolation is needed if one wants to use those results as a sub-grid model for large-scale CR diffusion. 

A possible role of anisotropic diffusion in large-scale CR transport modeling has been already mentioned in \cite{Evoli:2012ha}. In that paper, the tension between the observed CR gradient (with respect to the Galactocentric radius $R$) and the (steeper) one predicted by large-scale models is discussed. A viable solution is provided, motivated by the idea that CR {perpendicular} escape is more effective in highly star-forming regions where a larger turbulence level is expected (as shown in eq. \ref{eq:intro4}); the same idea was shown to provide a solution to the long-standing CR anisotropy problem as well. However, a full anisotropic treatment was not actually implemented in that context.

In a broader perspective, recent analyses \cite{Pakmor2016} of Galaxy formation based on the moving mesh code {\tt AREPO}, which includes injection of CRs from supernovae and CR transport (either isotropic or anisotropic), show that the isotropic approximation can produce incorrect and unrealistic results, and in particular cannot account for the significant magnetic field amplification needed to reproduce current observations.
According to this scenario, anisotropic CR diffusion may be also responsible of the strong vertical winds which are observed in the inner Galaxy and may play a role in the interpretation of the {\it Fermi Bubbles}, huge structures in the gamma-ray sky (previously observed in the radio)  extending $\simeq 50^{\circ}$ above and below the Galactic center \cite{Bubbles2010}. 

An accurate modeling of anisotropic diffusion became further necessary after recent astronomical observations revealed the presence of a poloidal component in the galactic magnetic field.
In fact, several authors \cite{2012ApJ...757...14J, 2016A&A...600A..29T} have exploited the most recent synchrotron measurements and outlined the presence of a complex magnetic topology, characterized, in the Galactic bulge, by a poloidal structure extending significantly far away from the Galactic plane similarly to what observed in external spiral galaxies  \cite{2016A&ARv..24....4B}.
Due to the expected highly anisotropic transport regime, this structure should impact on CR transport{, as also highlighted by recent observations of the CR dipole anisotropy and phase \cite{SchwadronSCI2014,AhlersPRL2016}}.

Given all these considerations and results, we propose here the idea that the progressive hardening in the CR proton slope { towards the Galactic centre} can be interpreted as a signature of anisotropic diffusion, and provide a fully tested numerical implementation of this process (an alternative implementation of anisotropic diffusion, based on a stochastic differential equation approach, can be found in \cite{2012A&A...547A.120E}). In our scenario, the harder slope inferred by Fermi-LAT data in the inner Galaxy is due to the {\it parallel} diffusive escape along the poloidal component of the large-scale regular magnetic field, characterized by a larger normalization and a harder scaling with rigidity with respect to the perpendicular diffusion dominating CR transport in the more external regions.  This picture provides a novel alternative to other recently proposed interpretations of the CR spectral hardening in the central region of the Galaxy: In particular, in \cite{Recchia2016}, the hardening is explained in a non-linear CR propagation context, while in \cite{Wolfendale2013} the hardening finds an explanation in the framework of anomalous diffusion.
 
The paper is structured as follows: In section $2$ we present an extended  version of the {\tt DRAGON2} code which implements inhomogeneous 2D anisotropic diffusion, under the assumption of azimuthal symmetry. The corresponding numerical tests are described in the Appendices. In section 3 we implement a toy model for the regular Galactic magnetic field that captures the main features relevant for the problem under consideration, and show how the behavior of the CR proton slope as a function of Galactocentric radius depends on the free parameters involved in our model. In section 4 we present a more realistic model that reproduces the data inferred by the Fermi-LAT collaboration. In section 5 we discuss our results and present our conclusions.

\section{Our setup}

The {\tt DRAGON} code\footnote{www.dragonproject.org} \cite{Evoli:2008dv} { is an appropriate framework to bridge the gap between theory and experimental data: In fact,  it solves a very general version of the transport equation, and the main motivation of the project, since the very beginning, has been to investigate what we can learn from CR, radio and gamma-ray measurements about the physics behind the CR transport problem and the origin of CRs themselves.
In particular, its latest version currently under development ({\tt DRAGON2} ~\cite{Evoli2017}), features all the relevant astrophysical ingredients required for CR propagation, and includes a complete set of the most up-to-date models of the large-scale regular magnetic field.}

In this work, we extend the  {\tt DRAGON2} code to solve the { two-dimensional CR transport equation including a fully-anisotropic diffusion tensor (i.e., keeping the assumption of azimuthal symmetry but with orientation of the local magnetic field).}   

The simplified scenario presented here { only focuses on proton propagation:} It captures the main aspects of the idea we are considering, and allows to make quantitative predictions on the spatial variation of the protons slope. 

We summarize below the main ingredients:

\begin{itemize}

\item {\it Geometry}: We assume azimuthal symmetry, so CR particles diffuse in a ($R\,$,$\,z$) plane defined by the following boundaries: $R\,\in\,[0, R_{\rm max}]$ and $z\,\in\,[-H, +H]$. In all the following simulations, $R_{\rm max}=20$~kpc and a resolution $dR=dz=0.1$~kpc has been adopted, while two values of the halo size, $H=2$ and $4$~kpc, have been investigated. 

\item {\it Transport equation}: We consider the following generic anisotropic transport equation:

\begin{equation}\label{eq:diff_eq_1}
 \frac{\partial\,N}{\partial t}\,=\,
 \bnabla\cdot\left({\bf D}\cdot\bnabla N\right)\,+\,S\,=\,
 \frac{\partial}{\partial x_i}\left(D_{ij}\,\frac{\partial\,N}{\partial x_j}\right)\,+\,S\,,
\end{equation}

where $N$ is the CR density, $S$ is the source term, {\bf D} is the diffusion tensor defined in eq.~(\ref{eq:D_ij_1}). We refer to the Appendix \ref{sec:appendix_diffusion} for all the details, including the numerical implementation.

\item {\it Source term}: Regarding the source spatial distribution, we consider the usual parametrization taken from \cite{Lorimer2006}, based on pulsar catalogs:

\begin{equation}
S(R, z) =  \left(\frac{R}{R_{\odot}}\right)^a \exp\left(-\,b\frac{R - R_{\odot}}{R_{\odot}}\,-\,\frac{\lvert z \rvert}{z_{0}}\right)\,,
\end{equation}
with $a=1.9$, $b=5$, $R_\odot=8.3$~kpc and $z_0=0.2$~kpc.

\item {\it Diffusion tensor}: 
Given a topology of the regular magnetic field, the diffusion tensor is naturally decomposed in the following way:

\begin{equation}\label{eq:D_ij_1}
D_{ij}\,\equiv\,D_\perp\delta_{ij}\,+\,\big(D_\|-D_\perp\big)b_ib_j\,,\qquad\,b_i\,\equiv\,\frac{B_i}{|\Bv|}\,,
\end{equation}

where $\Bv$ is the ordered magnetic field and $\bbv=\Bv/|\Bv|$ is its unit vector.
{ Note that in a complex magnetic configuration, the field orientation may vary in space, thus introducing a spatial dependence of the {\em diffusion tensor elements} $D_{ij}=D_{ij}(R,z)$ through the magnetic unit vectors, $b_i=b_i(R,z)$, without the necessity to assume spatially dependent {\em diffusion coefficients}, $D_\|$ and $D_\perp$. In fact, the parallel and perpendicular diffusion coefficients may reasonably depend only on the microphysics of the (axisymmetric) diffusion in a locally field-aligned coordinate system, which is encoded in their scaling with respect to the rigidity, and not necessarily on the spatial position itself.}
In this work we { indeed} consider spatially uniform and homogeneous { $D_\|$ and $D_\perp$}, but with different scalings with respect to the rigidity:
\begin{equation}\label{eq:DparaDperp_scalings-def}
D_\|\,=\,D_{0\|}\left(\frac{p_{\rm GeV}}{Z}\right)^{\delta_\|}\,\quad{\rm and}
\quad\,D_\perp\,=\,D_{0\perp}\left(\frac{p_{\rm GeV}}{Z}\right)^{\delta_\perp}\,\equiv\,
\epsilon_D\,D_{0\|}\left(\frac{p_{\rm GeV}}{Z}\right)^{\delta_\perp}\,,
\end{equation}
{ where $p_{\rm GeV}\equiv p/{\rm GeV}$ and, accordingly,} $\epsilon_D=D_{0\perp}/D_{0\|}$ is a parameter that gives the relative strength of the perpendicular to parallel diffusion coefficients at { the} reference energy { of} 1 GeV.

We remind the reader that, although in our setup $D_\|$ and $D_\perp$ are uniform, a spatially-dependent diffusion naturally arises from the dependence of the regular magnetic field orientation on the spatial coordinates as expressed by the unit vectors $b_R(R,z)$ and $b_z(R,z)$: Therefore, an effective phenomenological scenario of the type prescribed in Ref.~\cite{Gaggero:2014xla}, where a radially-dependent scaling $\delta(R)$ was introduced, naturally emerges within the framework of an anisotropic diffusion in the inhomogeneous Galactic magnetic field. 

\item {\it Magnetic field:} 
There has been a significant improvement in the modeling of the Galactic magnetic field (GMF) in the latest years. It is possible to identify a {\it regular component}, of a few $\mu G$ intensity, and coherent over $\mathcal{O}(\mathrm{kpc})$ lengths, which follows the large-scale structure of the Galaxy. The regular component can be modeled exploiting a wide set of data, including Faraday rotation measures and polarized synchrotron radiation, once a model for the CR electron distribution is provided.
Within the regular component, three relevant contributions are usually identified: the {\it disk}, the {\it halo}, and the {\it poloidal} magnetic fields. 

Given our 2D setup, we consider an azimuthal disk and halo component, with the intensity given by the following functions of $R$ and $z$, as in \cite{2011ApJ...738..192P}:

\begin{equation}\label{eq:Bphi_Bdisk}
B_{\phi}^{\rm disk} \, (R,z) \,= \, \left\{
 \begin{array}{cc}
  B_{D0} \, e^{-\lvert z \lvert/z_0} \,\,\,&  (R < R_{cD}) \\
  \\
  B_{D0} \, e^{-\lvert z \lvert/z_0} \, e^{-(R - R_{\circ})/R_0} \,\,\,&  (R > R_{cD}) 
 \end{array}
 \right.\,,
\end{equation}

\begin{equation}\label{eq:Bphi_Bhalo}
B_{\phi}^{\rm halo} \, (R,z) \,= \, B_{H0} \left[ 1 + \left( \frac{\lvert z \lvert  - z_0^H}{z_1^H} \right) \right]^{-1} \,\frac{R}{R_O^H} \, e^{\left(1 - \frac{R}{R_0^H}\right)}
\end{equation}
{ Note that, even in our 2D setup, it is important to model the $B_\phi$ component, although it enters the magnetic unit vectors $b_i$ only through the field strength, $|{\bf B}|=\sqrt{B_z^2+B_R^2+B_\phi^2}$. In fact, as we explicitly show in section \ref{sec:parametric} for a toy-model case, this component contributes to set a scale for the relative importance of parallel and perpendicular diffusion in Eq.~(\ref{eq:D_ij_1}).}

As far as the {\it poloidal} magnetic field is concerned, we adopt the parametrization given in \cite{2012ApJ...757...14J}, i.e., an X-shaped component with constant elevation from the Galactic disk for $R>R_{\rm X}^c$ and progressively more and more perpendicular to the Galactic plane as $R$ decreases below $R_{\rm X}^c$:
\begin{equation}\label{eq:BXX_Bz}
B_z^{\rm pol} \, (R,z) \,= \, B_{\rm X}(R,z)\,\cos\big[\Theta_{\rm X}(R,z)\big]\,,
\end{equation}
\begin{equation}\label{eq:BXX_BR}
B_R^{\rm pol} \, (R,z) \,= \, B_{\rm X}(R,z)\,\sin\big[\Theta_{\rm X}(R,z)\big]\,,
\end{equation}
with $B_{\rm X}$ and $\Theta_{\rm X}$ defined as
\begin{equation}\label{eq:BXX_bX}
B_{\rm X}(R,z) \,= \, \left\{
 \begin{array}{cc}
  B_{\rm X}^0 \, \left(\frac{R_p}{R}\right)^2 e^{-R_p/R_{\rm X}} \,\,\,&  (R \leq R_{\rm X}^c) \\
  \\
  B_{\rm X}^0 \, \left(\frac{R_p}{R}\right) e^{-R_p/R_{\rm X}} \,\,\,&  (R > R_{\rm X}^c)
 \end{array}
 \right.\,,
\end{equation}
\begin{equation}\label{eq:BXX_ThetaX}
\Theta_{\rm X}(R,z) \,= \, \left\{
 \begin{array}{cc}
  \tan^{-1}\left(\frac{|z|}{R-R_p}\right) \,\,\,&  (R \leq R_{\rm X}^c) \\
  \\
  \Theta_{\rm X}^0 \,\,\,&  (R > R_{\rm X}^c)
 \end{array}
 \right.\,,
\end{equation}
and
\begin{equation}\label{eq:BXX_Rp}
R_p\,= \, \left\{
 \begin{array}{cc}
   \frac{RR_{\rm X}^c}{R_{\rm X}^c+|z|/\tan\Theta_{\rm X}^0}\,\,\,&  (R \leq R_{\rm X}^c) \\
  \\
  R-\frac{|z|}{\tan\Theta_{\rm X}^0}\,\,\,&  (R > R_{\rm X}^c)
 \end{array}
 \right.\,,
\end{equation}
where the best-fits of the four free parameters have been determined in Ref.~\cite{2012ApJ...757...14J} to be $B_{\rm X}^0=4.6$~$\mu$G, $\Theta_{\rm X}^0=49^{\circ}$, $R_{\rm X}^c=4.8$~kpc, and $R_{\rm X}=2.9$~kpc.
The role of this magnetic field component is crucial in our setup, since it determines the progressively more and more ``vertical escape'' (i.e., along $z$) of the CRs in the {\it parallel} direction as $R$ decreases. This feature will be indeed characterized by a harder scaling of the CR spectrum with rigidity as $R$ decreases.

In figure~\ref{fig:Bfield_1} we provide a three-dimensional visualization of the complete magnetic field model described by eqs.~(\ref{eq:Bphi_Bdisk})--(\ref{eq:BXX_Rp}). 

\begin{figure}[ht!]	\includegraphics[width=\columnwidth]{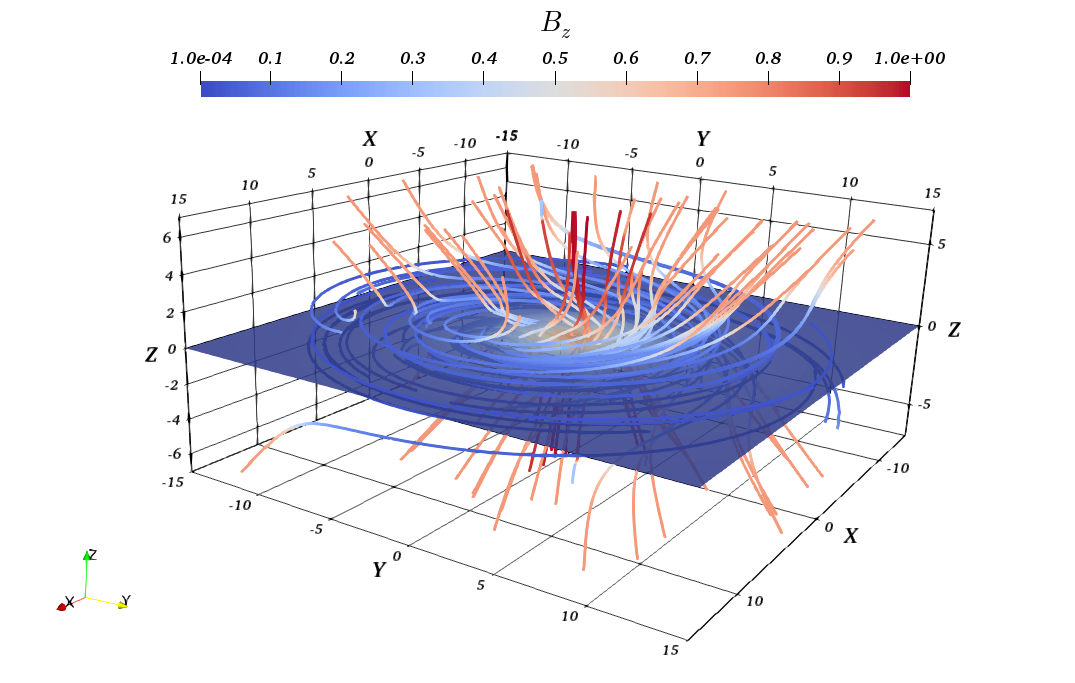}
    \caption{Three-dimensional representation of the Galactic regular magnetic field model used in our simulations and described by eqs.~(\ref{eq:Bphi_Bdisk})--(\ref{eq:BXX_Rp}). The values of the vertical component, $B_z$, is shown with colors on top of the magnetic field lines and as a contour plot on the $z=0$ Galactic plane. { Note that the field lines in the plot are randomly selected and the plot is meant for illustrative purpose only.}}
    \label{fig:Bfield_1}
\end{figure}

\item {\it Energy losses}: As far as hadronic particles are concerned, in the energy range we are considering the role of energy losses is negligible, as clearly shown e.g. in \cite{Bartels2017} (figure 1 in that paper): above 10 GeV the timescale for ionization and Coulomb energy losses exceed the diffusion timescales by at least 3 orders of magnitude.
\end{itemize}

\section{Parametric study}\label{sec:parametric}

As a preliminary step, we start by considering an even more simplified version of the magnetic field configuration presented above, in order to capture the main effects that we want to discuss and their dependence on the relevant parameters.
We therefore implement a toy-model magnetic field with no radial component, 
an exponentially decreasing component along $z$ and 
an exponentially increasing component along $\phi$ (i.e., out of the simulation plane):
\begin{align}
 B_R\,& =\,0\,\label{eq:toyBr}\\
 B_\phi\, & =\,B_{0,\phi}\left(1-e^{-R/R_0}\right)\label{eq:toyBphi}\\
 B_z\, & =\, B_{0,z}\,e^{-R/R_0} \,\equiv\, \epsilon_B\,B_{0,\phi}\,e^{-R/R_0}\,,\label{eq:toyBz}\,
\end{align}
where $\epsilon_B=B_{0,z}/B_{0, \phi}$ and $R_0$ are parameters that control the relative strength between $B_z$ and $B_\phi$, and their characteristic decay/growth length scales, respectively.

Given this toy-model magnetic field, we thus expect that the parallel diffusion dominates at small Galactocentric radii. In fact, at $R\ll R_0$, the disk component of the magnetic field is negligible and $\Bv$ is mainly along $z$, $B_\phi\ll B_z$: In such region the diffusion tensor reads
\begin{equation}\label{eq:toyB_D_small-r}
 {\bf D} = \left(
 \begin{array}{cc}
  D_{rr} & D_{rz}\\
  D_{zr} & D_{zz}
 \end{array}
 \right)\,\,\approx\,\left(
 \begin{array}{cc}
  D_\perp & 0\\
  0 & D_\|
 \end{array}
 \right)\,\approx\,D_\|\left(
 \begin{array}{cc}
  \epsilon_D\,p^{\delta_\perp-\delta_\|} & 0\\
  0 & 1
 \end{array}
 \right)\,\qquad\,R\ll R_0\,,
\end{equation}
so, if $\epsilon_D\ll1$ the cosmic rays perform just a relatively fast diffusion along $z$ (with minor modifications due to the perpendicular counterpart).
Conversely, at large $R\gg R_0$, the largest contribution to the magnetic field comes from $B_\phi\gg B_z$, which is the out-of-simulation-plane component, 
and therefore the diffusion becomes nearly isotropic:
\begin{equation}\label{eq:toyB_D_large-r}
 {\bf D}\,\approx\,\left(
 \begin{array}{cc}
  D_\perp & 0\\
  0 & D_\perp
 \end{array}
 \right)\,\qquad\,R\gg R_0\,,
\end{equation}
so that the cosmic rays are subject to a slower isotropic diffusion.

We expect that all these features will be reflected in a space-dependent spectral index of the CRs. 
By modeling the parallel and perpendicular diffusion coefficients as in eq.~(\ref{eq:DparaDperp_scalings-def}) and injecting the CRs with a power law of the type $S\propto p^{-\alpha_{\rm inj}}$, when $\epsilon_D\ll 1$ holds we expect that the stationary distribution of CRs in this magnetic field will be given by
\begin{equation}\label{eq:toyB_N_small-epsilonD}
 N(p)\,\approx\,\left\{
 \begin{array}{cc}
  p^{-(\alpha_{\rm inj}+\delta_\|)} & \qquad (R\ll R_0)\\
  &\\
  p^{-(\alpha_{\rm inj}+\delta_\perp)} & \qquad (R\gg R_0)\
 \end{array}
 \right.\,,
\end{equation}
i.e., in general, by a space-dependent power spectrum 
with a spectral index $\alpha(R)$ that depends on the distance from the galactic center.

{ It is important to remark that the picture depicted above could be altered in the $R \rightarrow 0$ region in the case where the ratio between the parallel and perpendicular diffusion coefficient is very large (e.g. when $\epsilon_D = 0.01$ and $\epsilon_B \ge 0.5$). Under these conditions, most of the CRs that are injected directly in the inner Galactic plane quickly diffuse in the vertical direction (parallel to the field) and leave the diffusive halo of the Galaxy. The proton spectrum in this central region receives therefore a dominant contribution from the particles that slowly diffuse perpendicularly along $R$ from larger radii. Given the steep rigidity dependence of the perpendicular diffusion coefficient, this contribution is more relevant at larger energies, thus determining a hardening that could be stronger than expected on the basis of the considerations exposed above (as it will be shown, in some limited cases, the spectral index could also be harder than the injection index).}

\begin{figure}[!t]
	\includegraphics[width=0.55\columnwidth]{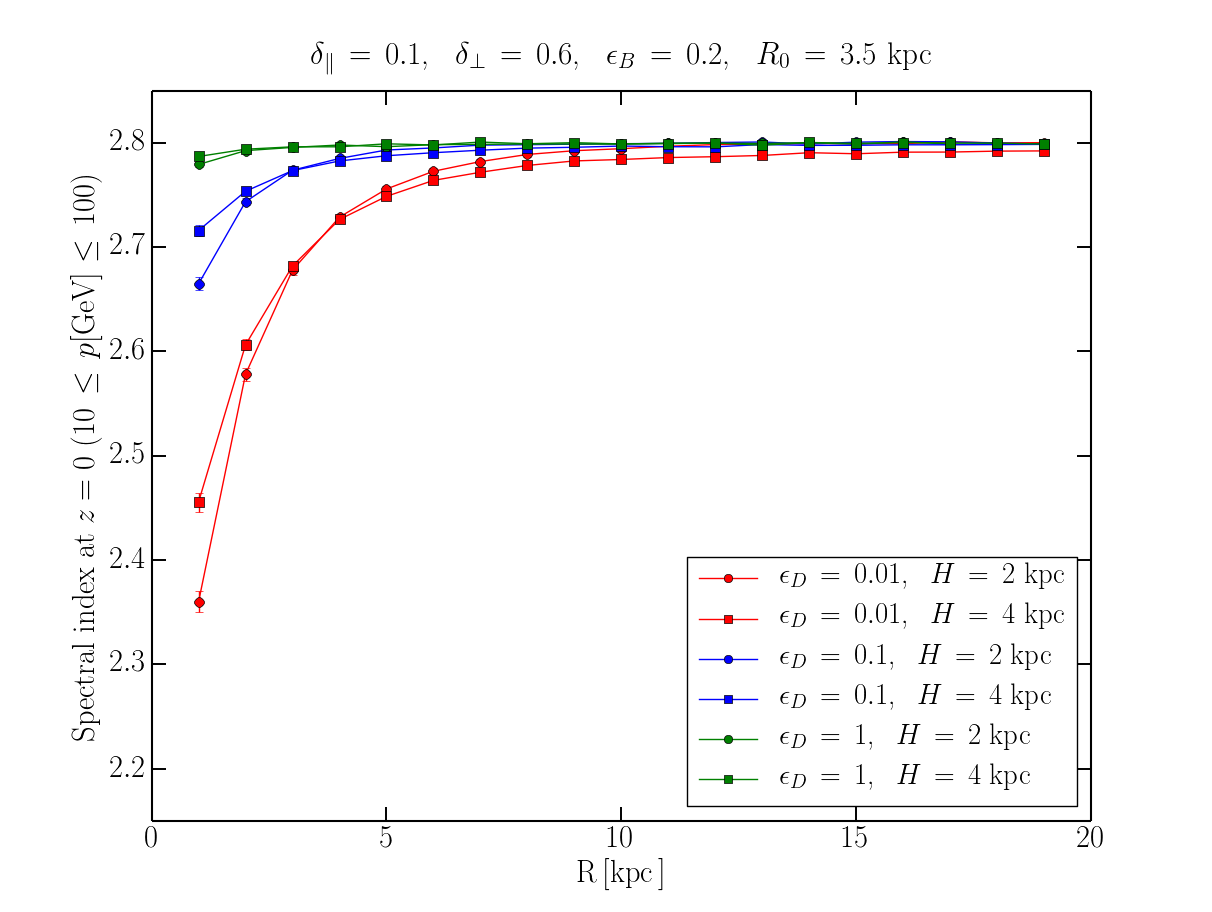}
	\hspace{-0.5cm}\includegraphics[width=0.55\columnwidth]{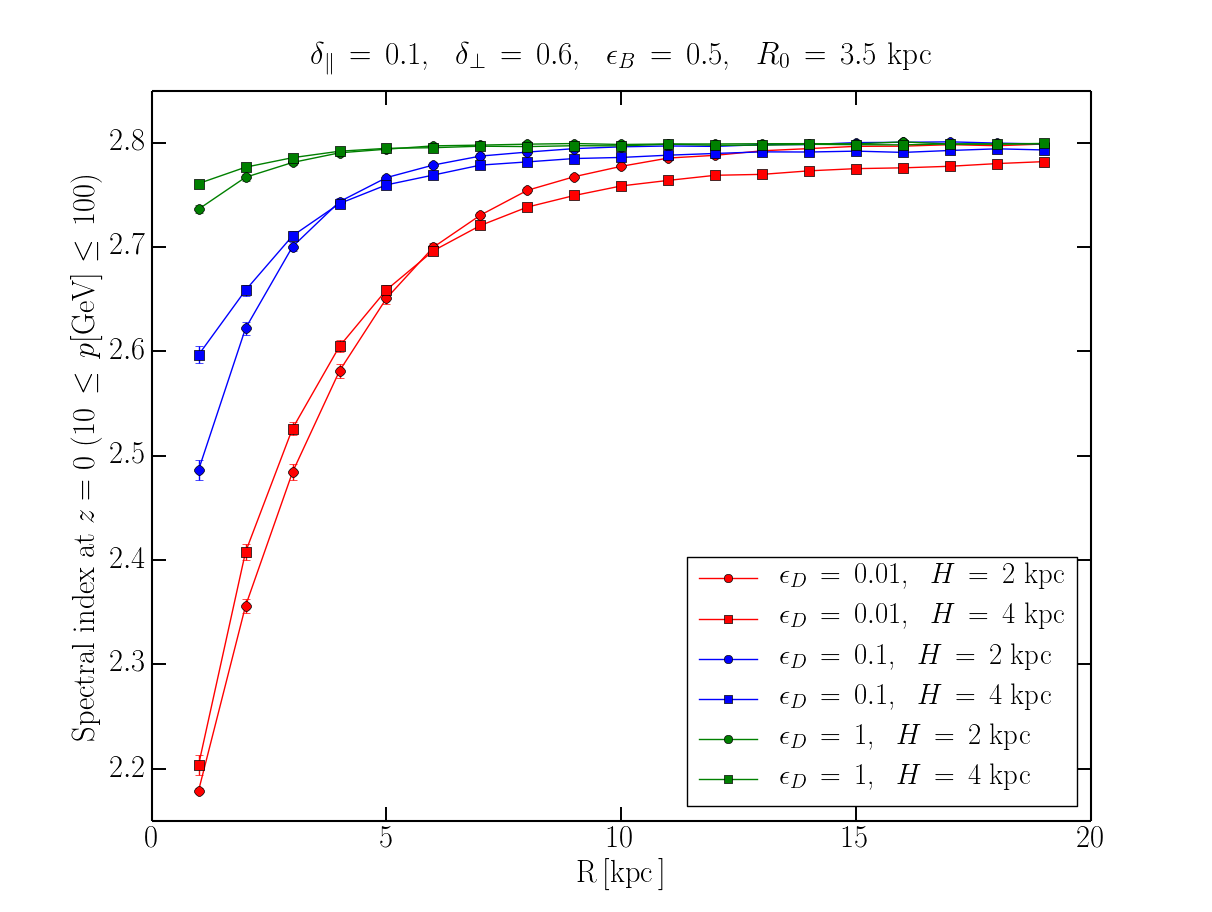}\\
	\includegraphics[width=0.55\columnwidth]{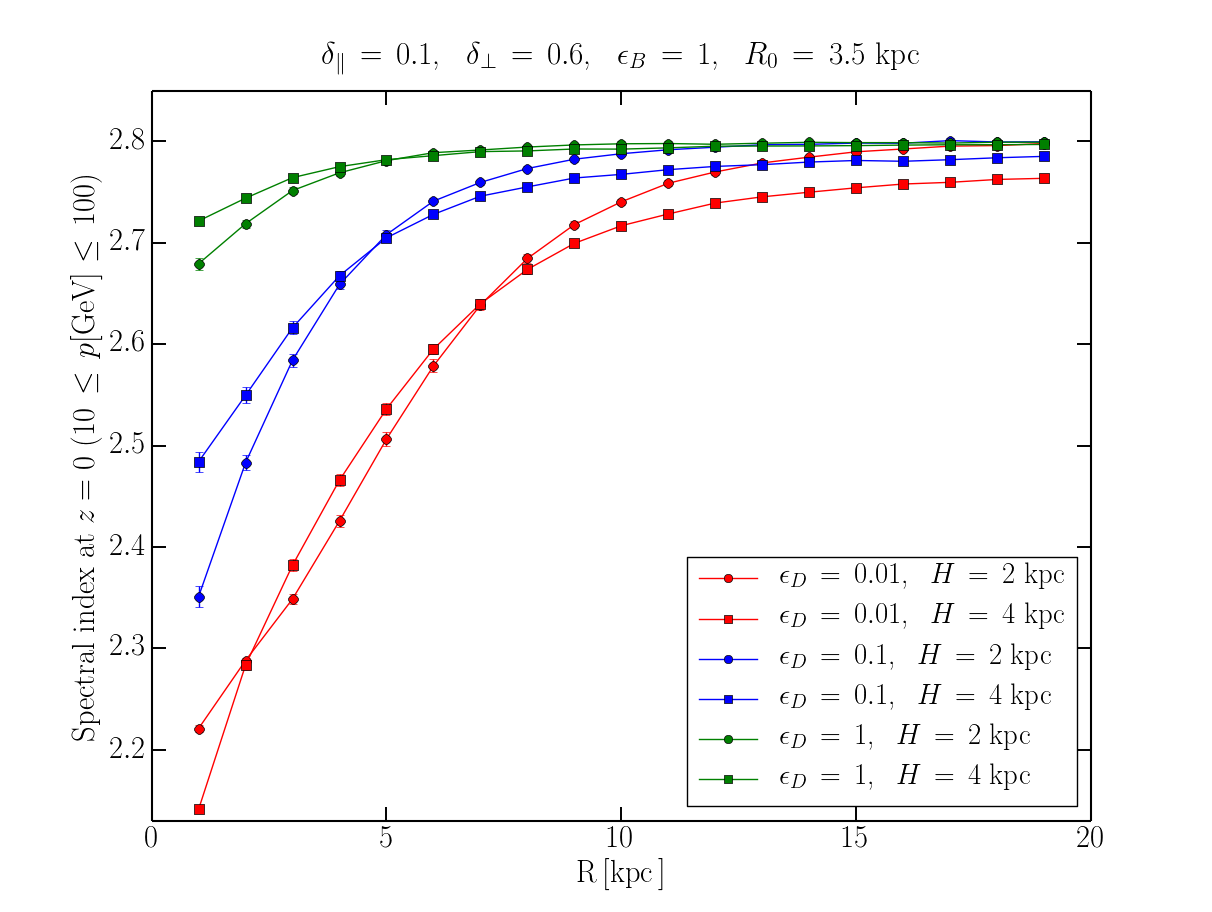}
	\hspace{-0.5cm}\includegraphics[width=0.55\columnwidth]{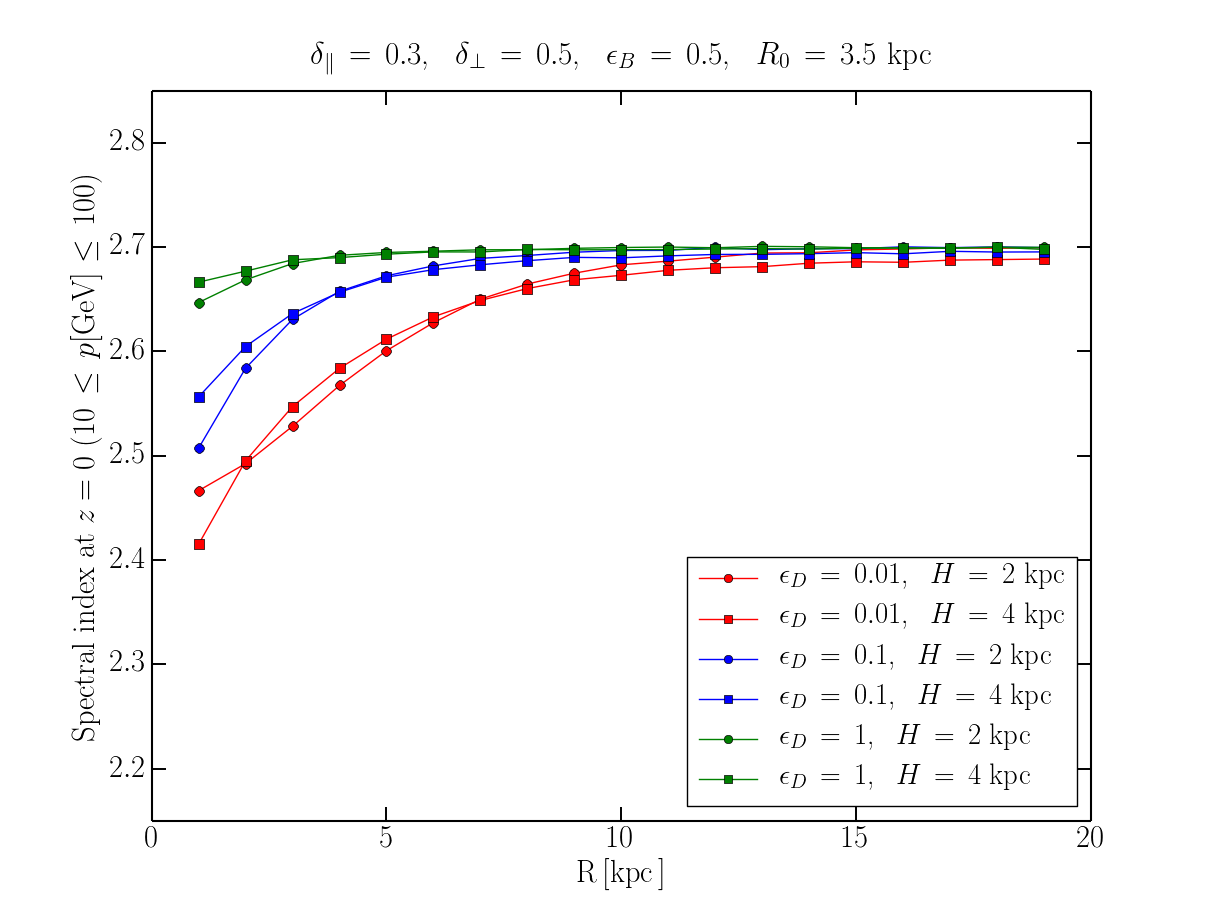}
    \caption{Fitted CR spectral index $\alpha$ as a function of the Galactocentric radius $R$ as the result of anisotropic diffusion within the toy-model magnetic field in eqs.~(\ref{eq:toyBr})--(\ref{eq:toyBz}) with $R_0=3.5$~kpc. { The spectral index at the injection is $\alpha_{\rm inj}=2.2$.} Different combinations of the relevant parameters, $\epsilon_B$, $\delta_\|$, $\delta_\perp$, $\epsilon_D$ and $H$, are reported (see title and legend in each panel).}
    \label{fig:parametric_main}
\end{figure}

In order to investigate the properties of the space-dependent anisotropic diffusion 
in the toy-model magnetic field described in eqs.~(\ref{eq:toyBr})--(\ref{eq:toyBz}), a number of simulations have been performed by varying several relevant parameters. 
In particular, the role of $\epsilon_B$, $\epsilon_D$, $\delta_\|$ and $\delta_\perp$ in determining $\alpha(R)$ has been explored taking into account also the possible effects of the halo size, $H$. 
Since scanning all the possible combinations of parameters would take unreasonably long time and resources, only few representative combinations have been considered here. 
Three values of $\epsilon_B=0.2$, $0.5$ and $1$ have been investigated 
for extremely anisotropic scalings of the diffusion coefficients, namely $\delta_\|=0.1$ and $\delta_\perp=0.6$ 
(we stress that this particular choice is not necessarily representative of a physical situation, 
but it is just to highlight the difference between the regions where one kind of diffusion dominates over the other).
A more realistic case with $\delta_\|=0.3$ and $\delta_\perp=0.5$ has also been considered 
for the intermediate case with $\epsilon_B=0.5$.
In all the cases mentioned above, only $R_0=3.5$~kpc has been considered 
(being a value close to the length scale of the poloidal magnetic field adopted for the more realistic GMF),
and two values of the halo size, $H=2$ and $4$ kpc, have been investigated. 
The results of this parametric study is reported in figure~\ref{fig:parametric_main}, 
where the fitted CR spectral index $\alpha$ as a function of the Galactocentric radius $R$ 
is shown for different values of $\epsilon_D$ and $H$ (see legend). 
 {Here we assume $\alpha_{\rm inj} = 2.2$.}
At large $R$, all the cases converge towards a nearly isotropic diffusion with $\alpha\simeq\alpha_{\rm inj}+\delta_\perp$ regardless of the $\epsilon_D$ value, whereas the spectral index at small $R$ is strongly dependent on such parameter (being $\alpha\simeq\alpha_{\rm inj}+\delta_\|$ when $\epsilon_D\ll1$, and getting closer to $\alpha\simeq\alpha_{\rm inj}+\delta_\perp$ as $\epsilon_D$ increases to unity). All the results are only weakly dependent on the values of $H$ that have been investigated. The behavior of the CR spectral index $\alpha$ in this toy-model magnetic field can thus be summarized as follows: 
\begin{itemize}
\item[(i)] The $\epsilon_D$ parameter controls the minimum and maximum values, $\alpha_{\rm min}$ and $\alpha_{\rm max}$, that can be reached by the CR spectral index $\alpha$ (at $R\ll R_0$ and at $R\gg R_0$, respectively, for the magnetic configuration considered here). As soon as $\epsilon_D\ll1$ and $\delta_\|<\delta_\perp$, the minimum value is $\alpha_{\rm min}\simeq\alpha_{\rm inj}+\delta_\|$ and increases to somewhat close to (but smaller than) $\alpha_{\rm inj}+\delta_\perp$ as $\epsilon_D\to1$, whereas $\alpha_{\rm max}\simeq\alpha_{\rm inj}+\delta_\perp$ regardless of $\epsilon_D$.
\item[(ii)] The length scale of the transition between $\alpha_{\rm min}$ and $\alpha_{\rm max}$ is essentially set by $R_0$, with some modification due to the parameter $\epsilon_B$, i.e. the length scale being slightly larger than just $R_0$ for increasing $\epsilon_B$.
\item[(iii)] The halo size $H$ weakly determines the relative weight of $\delta_\perp$ and $\delta_\|$ for small $R$, where the parallel diffusion along $z$ is supposed to dominate and the CRs are confined only for $|z|\leq H$. In fact, a quick escape in the $z$ direction can be partially counterbalanced by a very extended halo, thus allowing for additional perpendicular diffusion which then leaves a more effective fingerprint on the CR spectral index.
\end{itemize}

\section{Results for a realistic model}

Once the dependence of the proton slope on the relevant parameters of the toy model is understood, we consider the full implementation of the magnetic field described in Section 2.

Our aim is to present a set of reference propagation models characterized by a good agreement with: {\it 1)} the results of the model-independent analysis performed by the Fermi-LAT collaboration \cite{Acero2016ApJS}, which provides an estimate of the proton slope inferred by gamma-ray data at different Galactocentric radii; {\it 2)} the low-energy extrapolations of the numerical simulations presented in \cite{2002PhRvD..65b3002C, DeMarco2007a, Snodin2016}. In these studies, the propagation of charged test particles in a turbulent field is computed. In particular, the most recent one \cite{Snodin2016} covers a range of particle energies corresponding to $10^{-2} < R_L/\ell_c < 10^3$ (where $\ell_c$ is the magnetic correlation length\footnote{We remind the reader that { the energies covered by \cite{Snodin2016}} are at least $3$ orders of magnitude larger than those we are dealing with in the present work, so we have to assume that a low-energy extrapolation of the trends is possible}), and provides explicit expressions for the CR diffusion tensor: A clear evidence for a steeper scaling of the perpendicular diffusion coefficient is reported: $\delta_{\perp} \in [0.54, 0.72]$, in agreement with the earlier results discussed in \cite{DeMarco2007a}.

\begin{figure}[t]	\includegraphics[width=0.48\columnwidth]{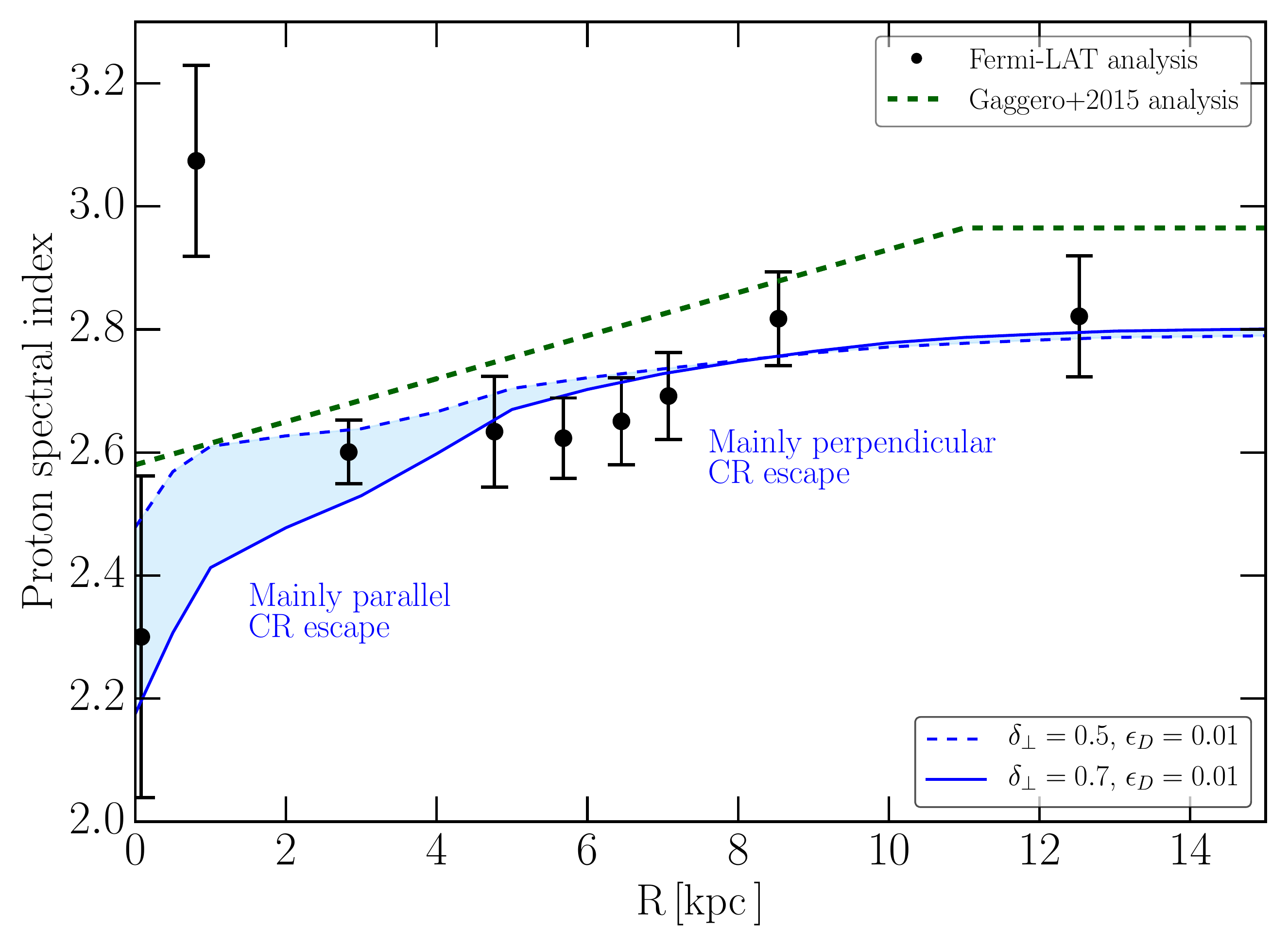}
\includegraphics[width=0.48\columnwidth]{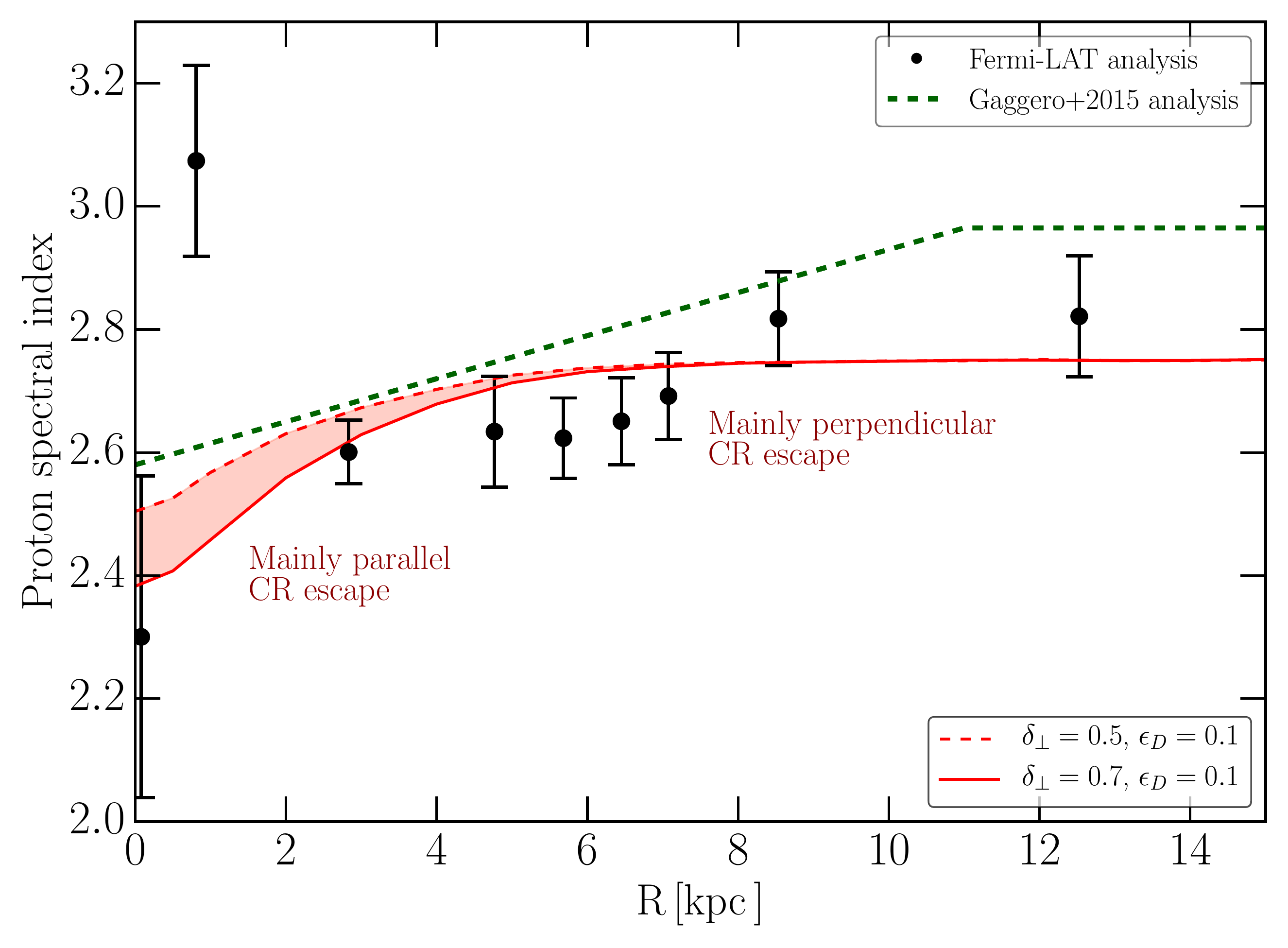}
    \caption{The predicted variation of the proton slope with respect to the Galactocentric radius is shown for two different choices of $\epsilon_D=D_{0\perp}/D_{0\|}$ ({\it left panel: $\epsilon_D = 0.01$}; {\it right panel: $\epsilon_D= 0.1$}). The model predictions are compared to the results of the gamma-ray analysis performed in \cite{Acero2016ApJS}; as a reference, we show the phenomenological model proposed in \cite{Gaggero:2014xla}.
	}
    \label{fig:mainPlot}
\end{figure}

Guided by those considerations we present a set of reference models: We adopt $\delta_{0\|} = 0.3$ and $\delta_{\perp} \in [0.5, 0.7]$, and consider two different scenarios characterized by $\epsilon_D  = 0.1$ and $\epsilon_D = 0.01$. The height of the diffusive halo is fixed to the commonly used benchmark value $H = 4$ kpc.  We show the main results in figure~\ref{fig:mainPlot}. 

We obtain in all cases a significant hardening of the proton spectrum. More precisely, the difference in slope between the local spectrum and the spectrum at the Galactic Center is [0.24, 0.36] if $\epsilon_D = 0.1$ or [0.27, 0.58] if $\epsilon_D = 0.01$, the lower bound of the interval corresponding to $\delta_\perp = 0.5$, while the upper one refers to the case where $\delta_\perp = 0.7$.
The hardening is compatible with the Fermi-LAT analysis, and in good agreement with the phenomenological interpretation based on a radially-dependent $\delta$ (in the context of isotropic diffusion) described in \cite{Gaggero:2014xla}.

It is very important to point out that in our results the hardening is always present in the energy range in which diffusion is dominant.
This differentiates our model from the scenario described in \cite{Recchia2016}, where the hardening is explained in the context of non-linear CR propagation, in which the transport is due to particle scattering -- and advection -- off turbulence generated by {\it streaming instability}, i.e. triggered by CRs themselves. The key idea of that explanation is that the diffusion coefficient is expected to be smaller in the inner regions -- closer to the peak of the source distribution -- due to the larger CR gradients (and, as a consequence, to the larger growth rate of CR-driven Alfv\'en waves): Therefore, the range in which advection dominates over diffusion, and the CR spectrum stays closer to the injection one, extends to larger momenta compared to the conventional scenario.
This effect can play indeed a role, especially at low energies; however, the authors point out that self-generating turbulence is relevant mostly below $\sim 50$ GeV. 
Given the results reported in \cite{Gaggero:2014xla}, in the inner Galaxy there is actually a hint of a harder gamma-ray spectrum dominated by the hadronic component (i.e., a harder proton spectrum compared to the local one) at least up to energies as large as $100$ GeV: A detailed model-independent analysis of Fermi-LAT data focused on this high-energy domain is therefore needed to  distinguish between the two scenarios. 

\begin{figure}[t]
\centering
\includegraphics[width=0.6\columnwidth]{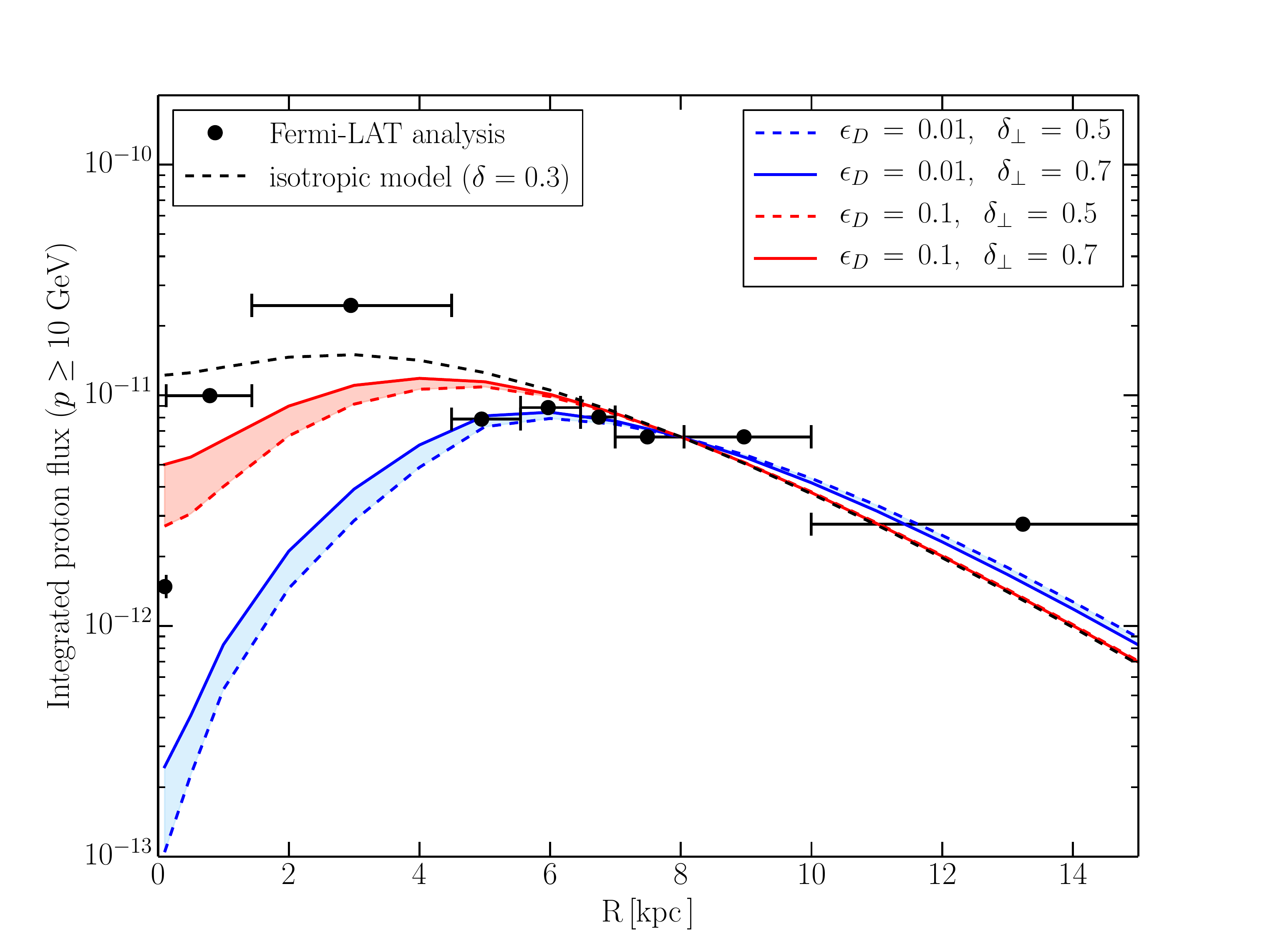}
    \caption{The impact of anisotropic diffusion on the CR flux on the Galactic plane at different Galactocentric radii is shown for the different setups considered in this work. As a comparison, we show the results of \cite{Acero2016ApJS}. A very high degree of anisotropy may result in a significant depletion of CRs in the inner Galaxy: See the detailed discussion in the text.
	}
    \label{fig:mainPlot_norm}
\end{figure}

Some comments about the impact of anisotropic diffusion on the CR density profile are now in order. 
For illustrative purposes we show in figure~\ref{fig:mainPlot_norm} the predicted profile of the CR density integrated above $p = 10$ GeV along the Galactic plane, compared to the results of the Fermi-LAT analysis discussed above \cite{Acero2016ApJS}.
In our setup, parallel diffusion in the inner Galaxy corresponds to an enhanced escape of particles in the vertical direction, given the topology of the magnetic field: Therefore, a high level of anisotropy necessarily implies a more efficient depletion of cosmic rays in the bulge. A similar trend can be qualitatively inferred from the first radial bins of the Fermi-LAT analysis, while{, in contrast with such analysis, the corresponding isotropic model predicts a negligibly small depletion of CRs in such region}.
On the other hand, we point out that the high CR density at $R \simeq 3$ kpc inferred by CR data and corresponding to the peak in the source distribution is highly smeared in our anisotropic scenarios. 
In this regard, we remark that the behavior of the injection term in the inner Galaxy is affected by extreme uncertainty (see the discussion in \cite{SpikePaper}), and inevitably a large degeneracy arises between the number and efficiency of CR sources in the central region and the effectiveness of parallel escape in the vertical direction. 
Moreover, it is appropriate to issue certain caveats of our 2D approach in this context: The magnetic field topology { is aligned with the spiral structure of the Galaxy}, hence we expect that the particles injected at $R \simeq$ 3 -- 4 kpc would propagate azimuthally along that pattern, thus possibly reducing, to a certain extent, the impact of the vertical escape.
In order to address these problems, a future development of this work is a study of the anisotropic transport in a fully 3D geometry.

{Finally, a comment is in order about the behavior of the proton spectral index at around 1 kpc. As clearly illustrated in Fig.~2, our model is not able to reproduce the very large fluctuation of the spectral index that is observed by Fermi-LAT at this Galactocentric distance. This could be a hint that other mechanisms might be competing with anisotropic diffusion in shaping the spectral index: It is interesting to notice, for example, that the very large softening of the spectrum at 1 kpc is reproduced well in the framework of the already discussed scenario described in Ref.~\cite{Recchia2016}, which, on the other hand, does not seem to explain the hardening that is observed at $R \rightarrow 0$ and that is correctly predicted by the model discussed here. In addition, one must consider also that the region at 1 kpc represents the outer Galactic bulge, a poorly understood region with very low gas density that may not be correctly captured by the Fermi-LAT gamma-ray analysis.}

\section{Conclusions}

The evidence for a progressive hardening of the proton spectrum towards the inner Galaxy represents a severe challenge for the standard, isotropic, homogeneous transport models considered so far in most of the literature.

In this work { we have explained this hardening in the context of anisotropic CR transport, and presented a complete description of (2D) numerical solver designed to compute the anisotropic diffusion of CR protons in the Galaxy within an azimuthally-averaged three-dimensional model for the Galactic magnetic field.}

At first, a toy-model magnetic field capturing the essential features of the realistic one has been considered, and the impact of the main free parameters on the CR propagated spectrum along the Galactic disk has been highlighted.

An azimuthally-symmetric version of the more recent and detailed models of the Galactic magnetic field has been then considered. Within such realistic implementation, we were able to reproduce the anomalous hardening of the CR proton spectral index recently inferred from the gamma-ray Fermi-LAT data

The results show for the first time a relevant observable signature of anisotropic diffusion and provides a consistent interpretation of a recently debated anomaly in the CR proton slope inferred by gamma-ray data. 

Further investigations of the anisotropic diffusion process in a full 3D setup, including other effects such as, e.g., the role of advection, will be needed in the future in order to capture an even wider phenomenology (e.g. the parallel escape along the spiral arms). 

\section*{Acknowledgements}

We are indebted to P.~Ullio and A.~Urbano for many interesting discussions and suggestions. We thank A.~Snodin for an interesting exchange about numerical simulations. S.~S.~Cerri gratefully acknowledges the GRAPPA institute for the hospitality during part of this project. A.~Vittino acknowledges support from the German-Israeli Foundation for Scientific Research.
\newpage

\appendix 

\section{The anisotropic diffusion equation}
\label{sec:appendix_diffusion}

The diffusion equation for CRs distribution $N(\xv,t)$ is 
\begin{equation}\label{app:diff_eq_1}
 \frac{\partial\,N}{\partial t}\,=\,
 \bnabla\cdot\left({\bf D}\cdot\bnabla N\right)\,+\,S\,=\,
 \frac{\partial}{\partial x_i}\left(D_{ij}\,\frac{\partial\,N}{\partial x_j}\right)\,+\,S\,,
\end{equation}
where we are dropping any species index and/or dependence for the sake of simplicity, $S$ represents a generic source term, and the elements of the {\em anisotropic} diffusion matrix can be written in general as
\begin{equation}\label{eq:D_ij}
 D_{ij}\,\equiv\,D_\perp\delta_{ij}\,+\,\big(D_\|-D_\perp\big)b_ib_j\,,\qquad\,b_i\,\equiv\,\frac{B_i}{|\Bv|}\,,
\end{equation}
where $\Bv$ is the ordered magnetic field, so $\bbv=\Bv/|\Bv|$ is the unit vector along it\footnote{Note that $\bnabla\cdot\Bv=0$, but $\bnabla\cdot\bbv=-\nabla_\|\ln(B)\neq0$ in general.}. The diffusion equation (\ref{app:diff_eq_1}) can also be rewritten in the following form:
\begin{equation}\label{eq:diff_eq_2}
 \frac{\partial\,N}{\partial t}\,=\,
 {\cal D}\big[N\big]\,+\,\uv_D\cdot\bnabla N\,+\,S\,,
\end{equation}
where ${\cal D}\equiv{\bf D}:\bnabla\bnabla=D_{ij}\partial_i\partial_j$ is a second-order differential operator (i.e., the usual diffusion operator $D\nabla^2$ for an isotropic and homogeneous diffusion tensor) and $\uv_D\equiv\bnabla\cdot{\bf D}$ is a drift-like velocity arising from the inhomogeneities present in the system (that are here seen through the diffusion coefficients). However, note that a drift-like term can arise not only for inhomogeneities, but also because of the form of the divergence in non-cartesian coordinate systems (see below for cylindrical coordinates). Hereafter, we are going to consider the two-dimensional (2D) case only, so, for instance, any dependence on $\phi$ in cylindrical coordinates will be neglected (azimuthally-symmetric approximation).

In the 2D approximation of cylindrical coordinates, $(R,z)$, 
the above equation can be rewritten as
\begin{equation}\label{eq:diff_eq_2Dcyl}
 \frac{\partial\,N}{\partial t}\,=\,
 D_{RR}\,\frac{\partial^2N}{\partial R^2}\,+\,D_{zz}\,\frac{\partial^2N}{\partial z^2}\,+\,
 2\,D_{Rz}\,\frac{\partial^2N}{\partial R\partial z}\,+\,
 u_R\,\frac{\partial\,N}{\partial R}\, +\, u_z\,\frac{\partial\,N}{\partial z}\,+\,S\,,
\end{equation}
where the drift-like velocities are given by
\begin{equation}\label{eq:drift-like_vel_2Dcyl}
 u_R\,=\,\frac{D_{RR}}{R}\,+\,\frac{\partial\,D_{RR}}{\partial R}\,+\,
 \frac{\partial\,D_{Rz}}{\partial z}\,\quad{\rm and}\quad\,
 u_z\,=\,\frac{D_{Rz}}{R}\,+\,\frac{\partial\,D_{Rz}}{\partial R}\,+\,
 \frac{\partial\,D_{zz}}{\partial z}\,.
\end{equation}

In the 2D cartesian coordinates $(x,y)$, eq.~(\ref{eq:diff_eq_2Dcyl}) retains the same form with just the coordinate change $r\to x$, but the drift-like velocities now read
\begin{equation}\label{eq:drift-like_vel_2Dcartesian}
 u_x\,=\,\frac{\partial\,D_{xx}}{\partial x}\,+\,
 \frac{\partial\,D_{xz}}{\partial z}\,\qquad{\rm and}\qquad\,
 u_z\,=\,\frac{\partial\,D_{xz}}{\partial x}\,+\,
 \frac{\partial\,D_{zz}}{\partial z}\,.
\end{equation}
By comparison of eqs.~(\ref{eq:drift-like_vel_2Dcartesian}) with eqs.~(\ref{eq:drift-like_vel_2Dcyl}) it is clear that even if the diffusion coefficients are constants, a ``metric-induced'' dirft may sill be present in cylindrical coordinates.

\subsection{Numerical implementation}

The diffusion equation (\ref{eq:diff_eq_2Dcyl}) has been implemented via the so-called operator splitting (OS) method. The right-hand-side operator has been splitted into three parts: one for the propagation in $R$, another one for the propagation in $z$ and a third part dealing with the mixed second-order derivative term. The first two operators are treated with the Crank-Nicolson (CN) scheme, i.e.,
\begin{eqnarray*}\label{eq:CNscheme_r}
 \left(1+\frac{\Delta t}{2}{\cal C}_{i,j}^R\right)N_{i,j}^{n+1}-\frac{\Delta t}{2}{\cal U}_{i,j}^RN_{i+1,j}^{n+1}
 -\frac{\Delta t}{2}{\cal L}_{i,j}^R N_{i-1,j}^{n+1}=\\
 = \, \left(1-\frac{\Delta t}{2}{\cal C}_{i,j}^R\right)N_{i,j}^n+\frac{\Delta t}{2}{\cal U}_{i,j}^R N_{i+1,j}^n
 +\frac{\Delta t}{2}{\cal L}_{i,j}^RN_{i-1,j}^n\,+\,\frac{\Delta t}{n_{\rm os}}\,S_{i,j}
\end{eqnarray*}
where $n_{os}=3$ is the number of splitting done with the OS method, ${\cal C}^R$, ${\cal U}^R$ and ${\cal L}^R$ are the central diagonal, upper diagonal and lower diagonal terms for the $R$-propagation operator, respectively, defined by
\begin{equation}\label{eq:CentralDiag_r}
 {\cal C}_{ij}^R\,\equiv\,\frac{2\,(D_{RR})_{ij}}{\Delta R_u\,\Delta R_d}\,,
\end{equation}
\begin{equation}\label{eq:UpperDiag_r}
 {\cal U}_{ij}^R\,\equiv\,\frac{(D_{RR})_{ij}}{\Delta R_c\,\Delta R_u}\,+\,\frac{(u_R)_{ij}}{2\,\Delta R_c}\,,
\end{equation}
\begin{equation}\label{eq:LowerDiag_r}
 {\cal L}_{ij}^R\,\equiv\,\frac{(D_{RR})_{ij}}{\Delta R_c\,\Delta R_d}\,-\,\frac{(u_R)_{ij}}{2\,\Delta R_c}\,,
\end{equation}
where $\Delta R_c\equiv(R_{i+1}-R_{i-1})/2$, $\Delta R_u\equiv R_{i+1}-R_i$, and $\Delta R_d\equiv R_i-R_{i-1}$. The propagation in $z$ has the same form (and the corresponding straightforward changes in the $i$ and $j$ indices), with the central, upper and lower diagonal terms given by the immediate symmetrization of the above:
\begin{equation}\label{eq:CentralDiag_z}
 {\cal C}_{ij}^Z\,\equiv\,\frac{2\,(D_{zz})_{ij}}{\Delta z_u\,\Delta z_d}\,,
\end{equation}
\begin{equation}\label{eq:UpperDiag_z}
 {\cal U}_{ij}^Z\,\equiv\,\frac{(D_{zz})_{ij}}{\Delta z_c\,\Delta z_u}\,+\,\frac{(u_z)_{ij}}{2\,\Delta z_c}\,,
\end{equation}
\begin{equation}\label{eq:LowerDiag_z}
 {\cal L}_{ij}^Z\,\equiv\,\frac{(D_{zz})_{ij}}{\Delta z_c\,\Delta z_d}\,-\,\frac{(u_z)_{ij}}{2\,\Delta z_c}\,.
\end{equation}
The third operator involving the mixed 2nd-order derivative term has been implemented with an explicit scheme:
\begin{equation}\label{eq:MixedDerivativeScheme}
 N_{i,j}^{n+1}\,=\,N_{i,j}^n\,+\,
 \Delta t\,{\cal M}_{i,j}\Big(N_{i+1,j+1}^n+N_{i+1,j-1}^n-N_{i-1,j+1}^n+N_{i-1,j-1}^n\Big)\,
 +\,\frac{\Delta t}{n_{\rm os}}\,S_{i,j}\,,
\end{equation}
with 
\begin{equation}\label{eq:MixedDiag_z}
 {\cal M}_{ij}\,\equiv\,\frac{(D_{Rz})_{ij}}{2\,\Delta R_c\,\Delta z_c}\,.
\end{equation}

\section{Numerical tests}

In the following, few basic simulations are presented in order to test the reliability of the numerical implementation and solution of eq.~(\ref{eq:diff_eq_2Dcyl}). These numerical tests are of two types:
\begin{itemize}
\item[(i)] Green function tests: Given an initial single point source localized in $\xv_{\rm src}$ and active at $t=0$ only, the solutions of the diffusion equation and their time evolution is known. In this test, the correctness of the time-dependent numerical solution with the corresponding analytical Green functions is verified (see Sec.~\ref{app:greenfunct}).
\item[(ii)] Comparisons with analytical solutions (following the approach of \cite{Kissmann:2014sia}): Given an expression for the stationary solution $N_{\rm an}(R,z)$, i.e., such that $\partial_tN=0$ and satisfying the correct boundary conditions, we can derive the analytical expression for the source $S_{\rm an}$ that is required. In this case, we test the convergence of the numerical solution, $N_{\rm num}$, to the expected analytical counterpart, $N_{\rm an}$, by employing the required source $S_{\rm an}$ (see Sec.~\ref{app:picardtest}).  
\end{itemize}

For all the tests presented here, we consider a simulation domain $R\in[0,R_{\rm max}]$ and $z\in[-H,+H]$ with $R_{\rm max}=2H=4$ kpc, discretized with $N_R=N_z=101$ uniformly distributed grid points for the Green function tests and with $N_R=N_z=31$ for the comparisons with analytical solutions.

\subsection{Green function}\label{app:greenfunct}

\begin{table}[t]
 \center
 \begin{tabular}{ccc|ccccccccccc}
  \toprule
    & Run & & & ($N_R$,$N_z$) & & $N_p$ & & $dt$ [Myr] & & ($b_R$,$b_z$) & & $D_{0\perp}/D_{0\|}$\\
   \hline 
    & G1 & & & (101,101) & & 2 & & 0.001 & & (0,1) & & 0.05 \\
    & G2 & & & (101,101) & & 2 & & 0.001 & & (1,1)/$\sqrt{2}$ & & 0.05 \\
    & G3 & & & (101,101) & & 2 & & 0.001 & & (1,1)/$\sqrt{2}$ & & 1 \\
   \hline
  \hline
 \end{tabular}
   \caption{Simulation parameters for Green-Function tests. The constants in physical units are $D_{0\|}=10^{28}$ cm$^2$/s and $B_0=10$ $\mu$G.
   }
   \label{tab:tab1}
\end{table}

\begin{figure}[h]
  \center
  \includegraphics[width=0.65\textwidth]{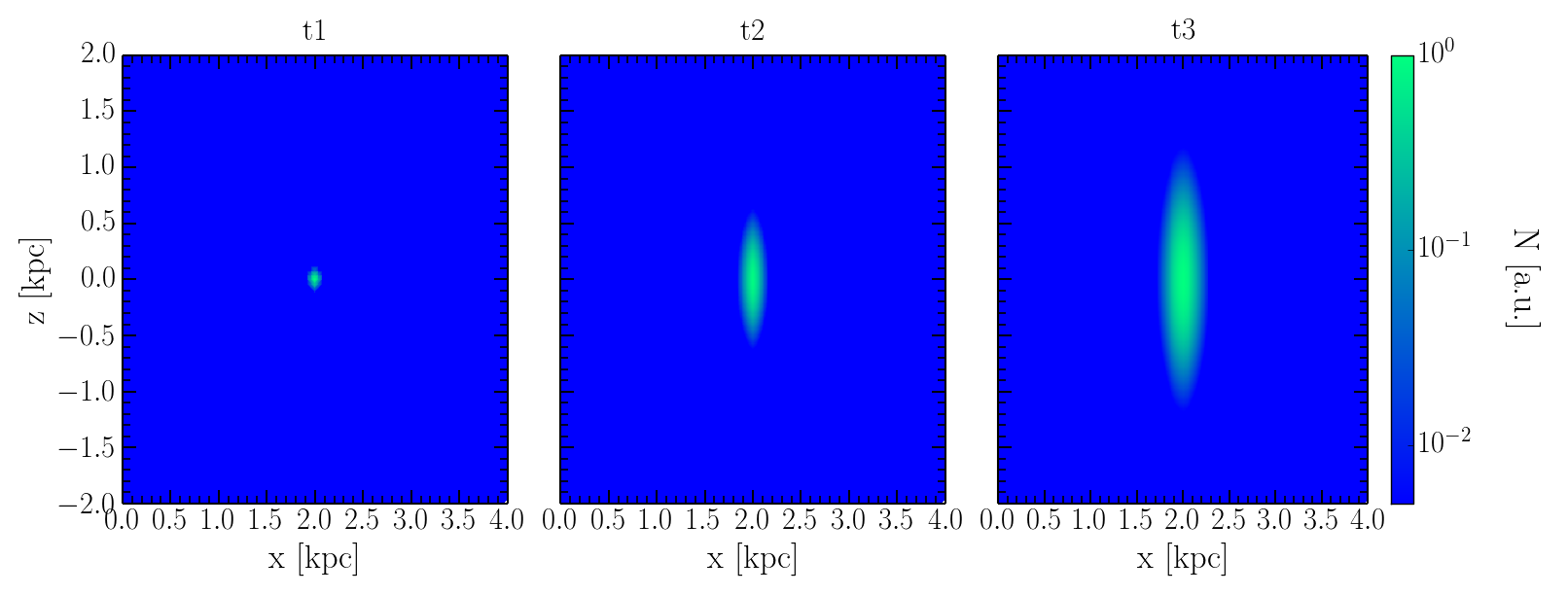}\\
  \includegraphics[width=0.35 \textwidth]{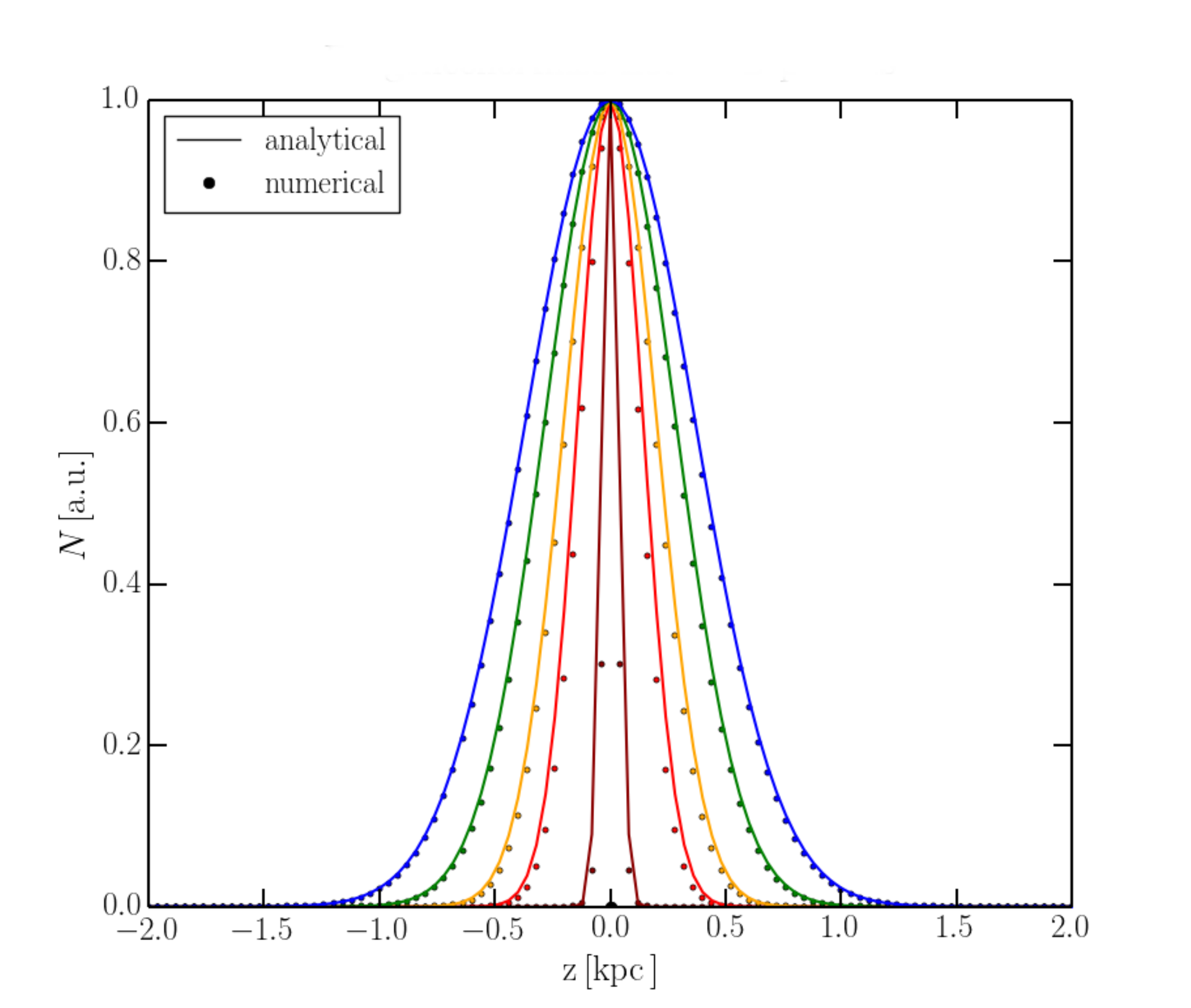}
  \includegraphics[width=0.35\textwidth]{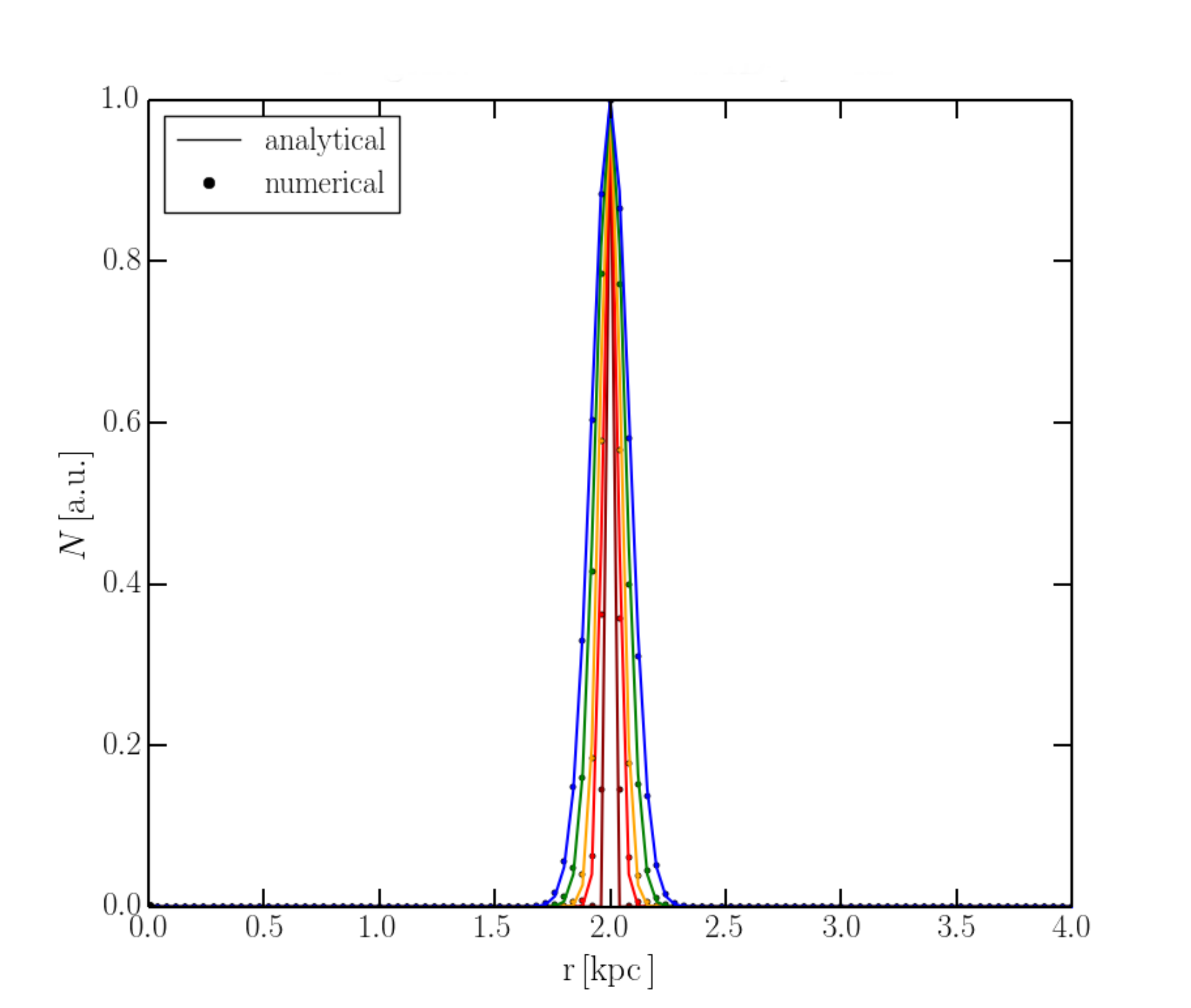}
 \caption{Upper panels: contour plots of $N(R,z,t)$ (normalized) at three different times.
  ~Lower panels: profiles $N(R_{\rm src},z,t)$ (left) and $N(R,z_{\rm src},t)$ (right) at different times (colors): comparison between numerical (circles) and analytical (solid lines) solution. Here $B_z=B_0$, $B_R=0$, $D_{0\perp}/D_{0\|}=0.05$, $N_R=N_z=101$ and $dt=0.001$ Myr.}
 \label{fig:test1a}
\end{figure}


\begin{figure}[h]
  \center
  \includegraphics[width=0.65\textwidth]{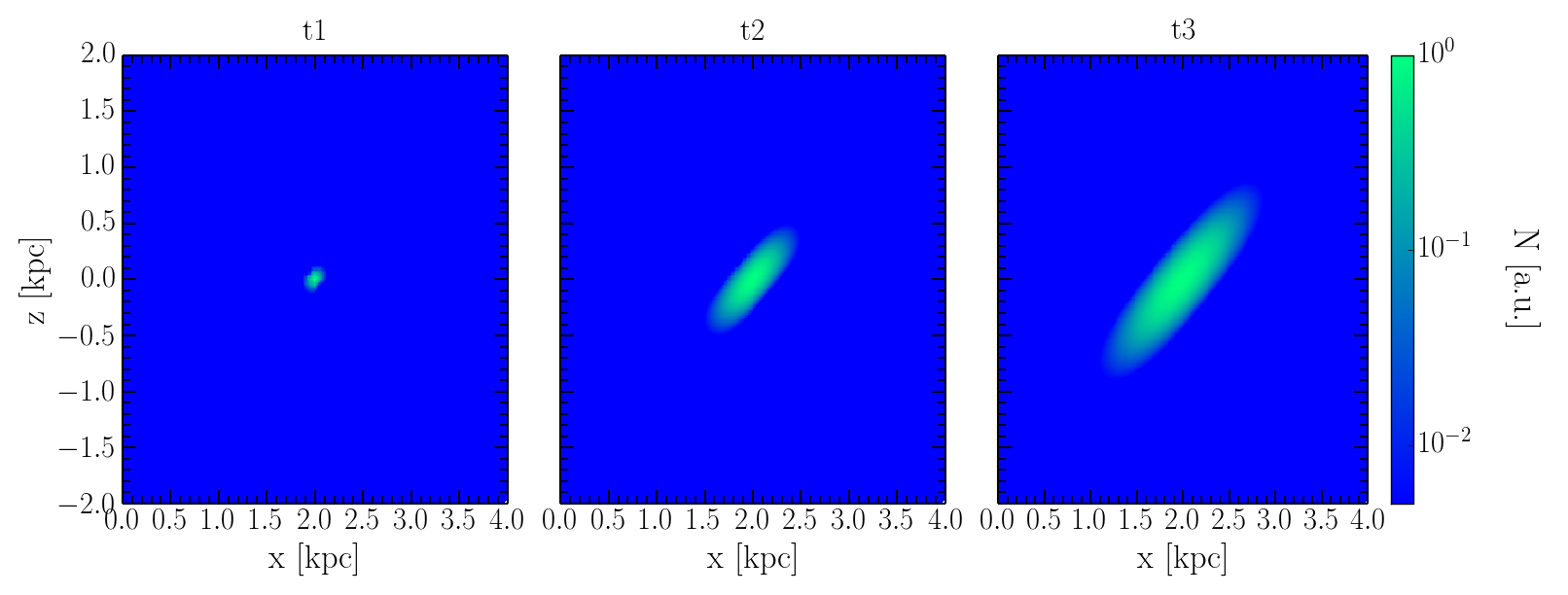}\\
  \includegraphics[width=0.35\textwidth]{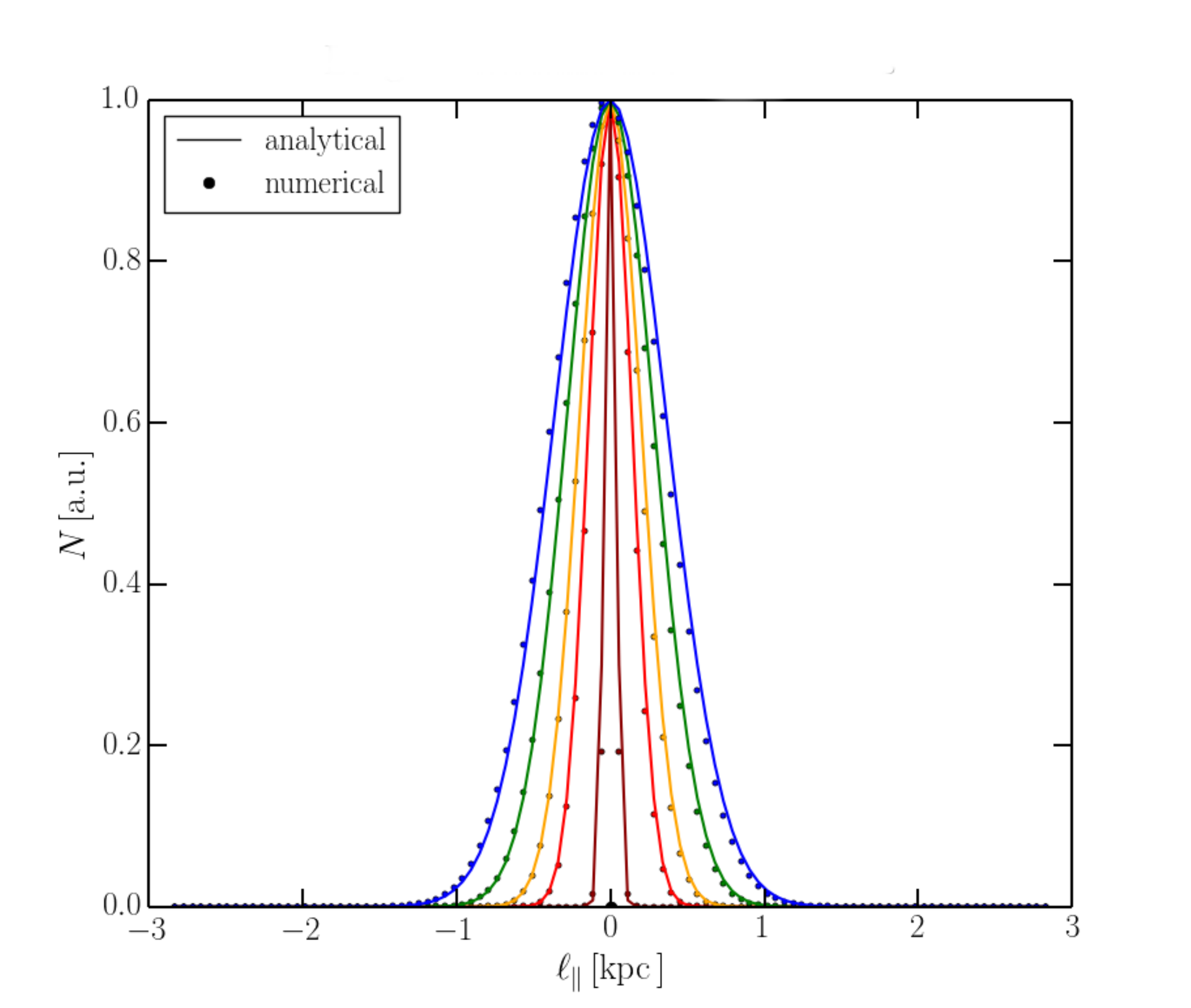}
  \includegraphics[width=0.35\textwidth]{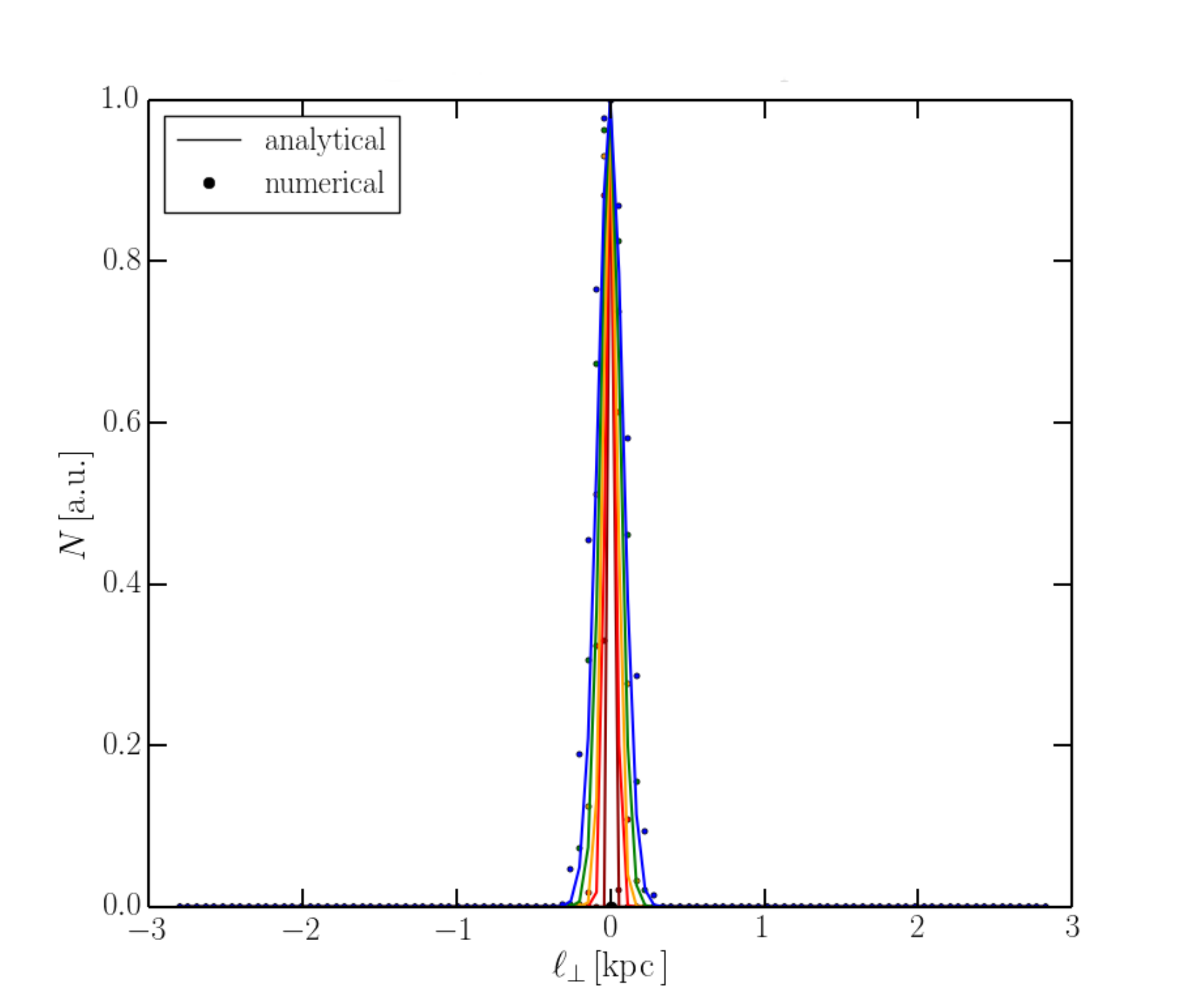}
 \caption{Upper panels: contour plots of $N(R,z,t)$ (normalized) at three different times.
 Lower panels: diagonal profiles, parallel (left) and perpendicular (right) to the magnetic field direction at different times (colors): comparison between numerical (circles) and analytical (solid lines) solution. Here $B_z=B_R=B_0$, $D_{0\perp}/D_{0\|}=0.05$, $N_R=N_z=101$ and $dt=0.001$ Myr.}
 \label{fig:test2a}
\end{figure} 

\begin{figure}[h]
  \center
  \includegraphics[width=0.65\textwidth]{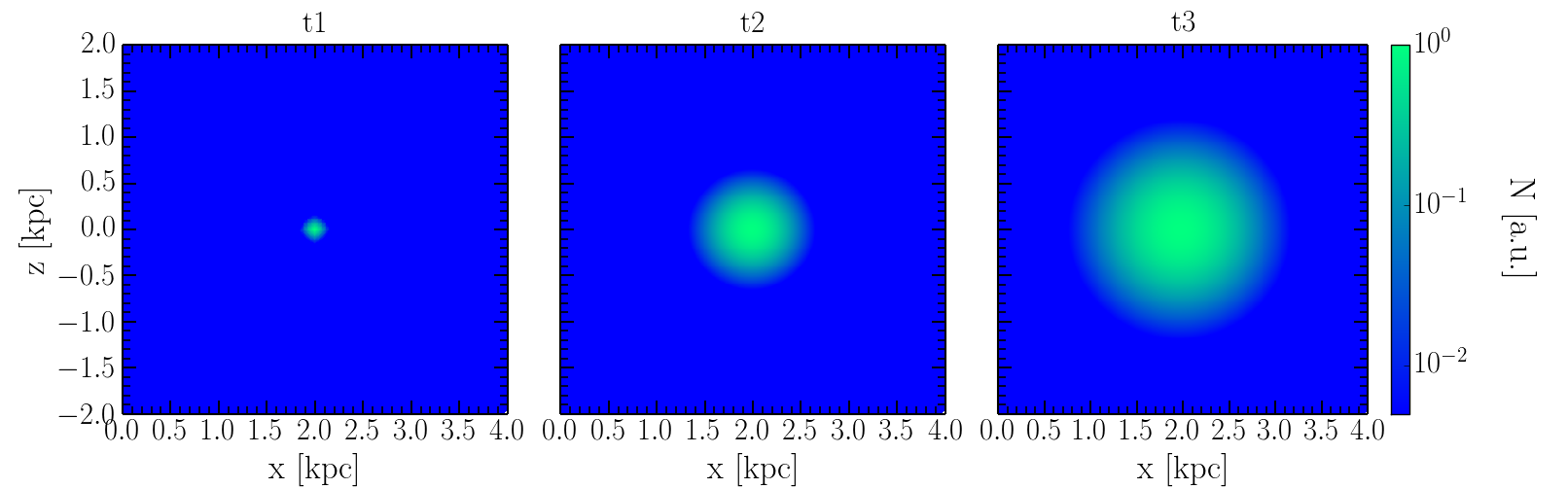}\\
  \includegraphics[width=0.35\textwidth]{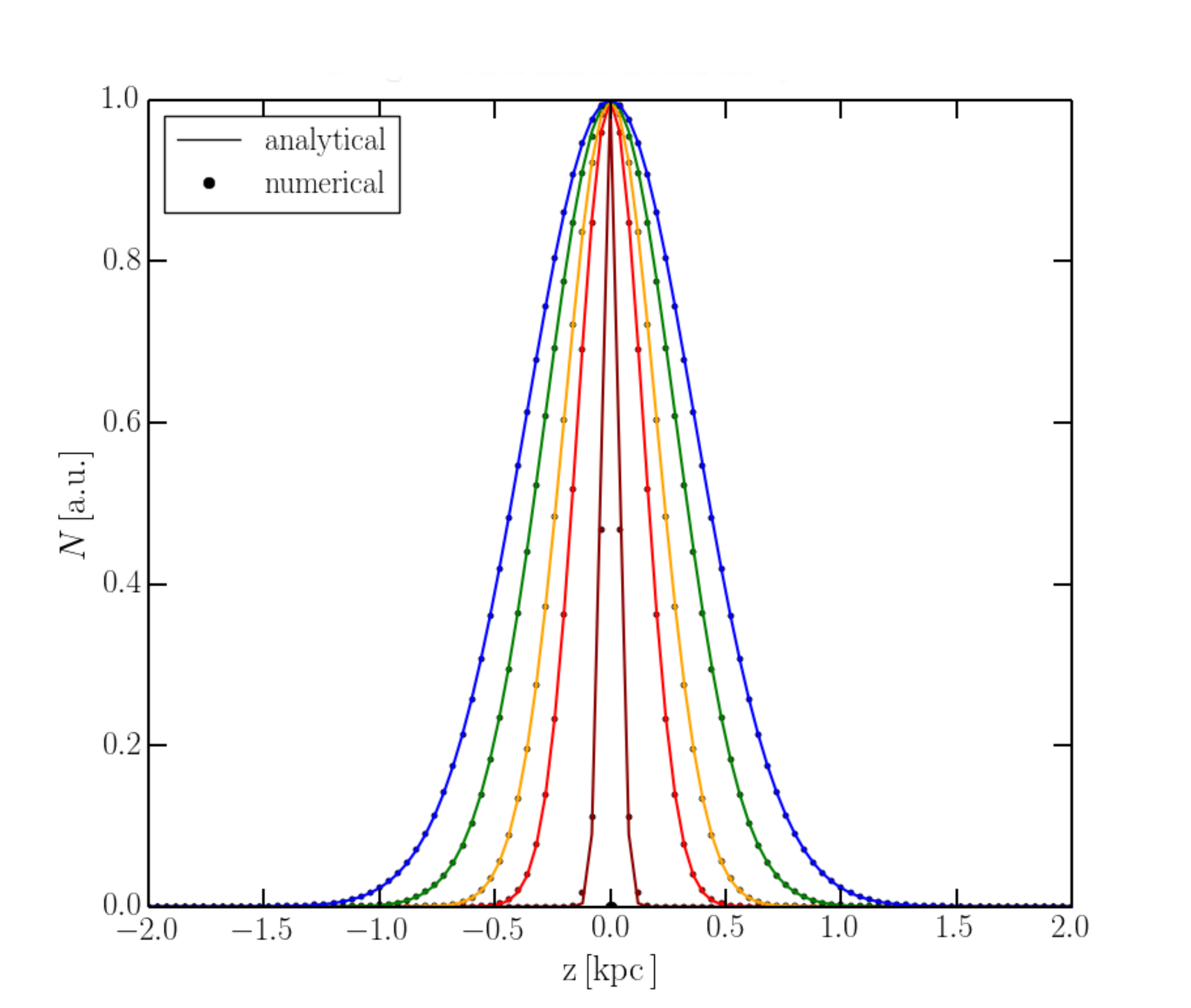}
  \includegraphics[width=0.35\textwidth]{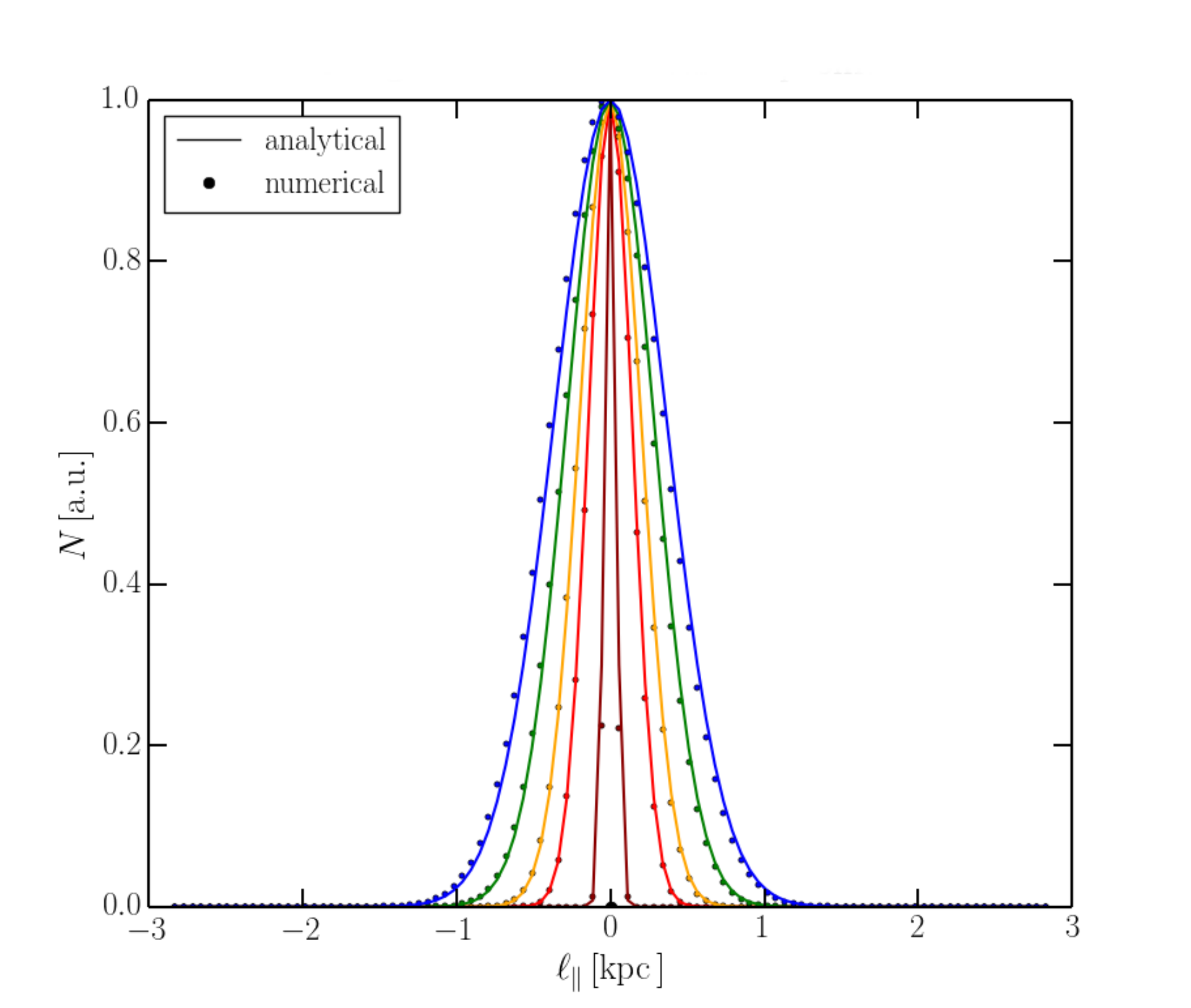}
 \caption{Upper panels: contour plots of $N(R,z,t)$ (normalized) at three different times.
 Lower panels: diagonal profiles, along $z$ (left) and parallel to the magnetic field direction (right) at different times (colors): comparison between numerical (circles) and analytical (solid lines) solution. Here $B_z=B_R=B_0$, $D_{0\perp}=D_{0\|}=D_0$, $N_R=N_z=101$ and $dt=0.001$ Myr.}
 \label{fig:test2c}
\end{figure}

We initialize the simulations with a point source located at $\xv_{\rm src}=(R_{\rm src},z_{\rm src})=(2,0)$ kpc, which is active only at $t=0$:
\begin{equation}\label{eq:PointSource_transient}
 S\,=\,\frac{\delta(t)}{\pi\Delta^2}\,e^{[(R-R_{\rm src})^2+(z-z_{\rm src})^2]/\Delta^2}\,,
\end{equation}
where $\Delta=\min(dR,dz)$. The solutions are the following well-known Green functions:
\begin{equation}\label{eq:GreenFunct_para}
 G_\|(\ell_\|,t)\,=\,\frac{1}{\sqrt{4D_\|t}}\,e^{l_\|^2/(4D_\|t)}\,,
\end{equation}
\begin{equation}\label{eq:GreenFunct_perp}
 G_\perp(\ell_\perp,t)\,=\,\frac{1}{\sqrt{4D_\perp t}}\,e^{l_\perp^2/(4D_\perp t)}\,,
\end{equation}
where $\ell_\|\equiv(\xv-\xv_{\rm src})\cdot\bbv$ is the direction parallel to the magnetic field and $\ell_\perp$ is the direction perpendicular to it.\\ 
In Table~\ref{tab:tab1}, we report the parameters used for the above tests. The results of such tests are shown in figures~\ref{fig:test1a}--\ref{fig:test2c}. Remarkable agreement has been found. Note that in the test G3, where $b_R=b_z\neq0$, the profiles are slightly shifted because drift terms are not zero, $u_R\neq0$ and $u_z\neq0$, even if the diffusion matrix elements are constants. Also, some minor discrepancies in that test may be due to the diagonal grid reconstruction. 

\subsection{Consistency with analytical solutions}\label{app:picardtest}


In this test, we define a priori the analytical form of a solution $\psi$ that we want to get in the stationary state (i.e., $\partial_t\psi=0$), accordingly to some boundary conditions. From such solution, we can explicitly derive the analytical form of the source term $S$ via the stationary diffusion equation, $S=-\bnabla\cdot({\bf D}\cdot\bnabla\psi)$.\\ 

In cylindrical coordinates, $(R,z)$, the boundary conditions (BCs) on a $[0,R_{\rm max}]\times[-H,+H]$ domain that have to be imposed on $\psi$ are the following:
\begin{align}\label{eq:BConPsi}
 \psi(z=\pm H)\,& =\,0\,,\\
 \psi(R=R_{\rm max})\, & =\, 0\,,\\
 \frac{\partial\,\psi}{\partial R}\bigg|_{R=0}\, & =\,0\,.
\end{align}
The simplest non-trivial function that fulfills the above BCs is given by
\begin{equation}\label{eq:PicardPsi}
 \psi\,=\,\psi_0\,\cos(k_RR)\cos(k_Hz)\,,
\end{equation}
where $k_R=\pi/2R_{\rm max}$ and $k_H=\pi/2H$, and $\psi_0$ is an arbitrary constant (hereafter, $\psi_0=1$). From $\psi$, we can derive the source term $S$ which would be necessary to produce such solution:
\begin{equation}\label{eq:PicardSource}
 S\,=\,-\bnabla\Big({\bf D}\cdot\bnabla\psi\Big)\,=\,
 \big(k_R^2D_{RR}+k_H^2D_{zz}\big)\psi\,-
 2\,D_{Rz}\,\frac{\partial^2\psi}{\partial R\partial z}\,
 -\, u_R\frac{\partial\,\psi}{\partial R}\,
 -\, u_z\frac{\partial\,\psi}{\partial z}\,,
\end{equation}
where $D_{ij}$, $u_R$ and $u_z$ are given in eqs.~(\ref{eq:D_ij}) and (\ref{eq:drift-like_vel_2Dcyl}). The derivatives of $\psi$ can be written down explicitly, so one gets
\begin{align}\label{eq:PicardSource_2}
 S\,= & \,\,\big(k_R^2D_{RR}+k_H^2D_{zz}\big)\cos(k_RR)\cos(k_Hz)\nonumber\\
      & \,-\,2\,k_R\,k_H\,D_{Rz}\,\sin(k_RR)\sin(k_Hz)\\
      & \,+\, k_R\,u_R\sin(k_RR)\cos(k_Hz)\,+\, k_H\,u_z\cos(k_RR)\sin(k_Hz)\,,\nonumber
\end{align}



\begin{figure}[h]
  \center
  \includegraphics[width=0.45\textwidth]{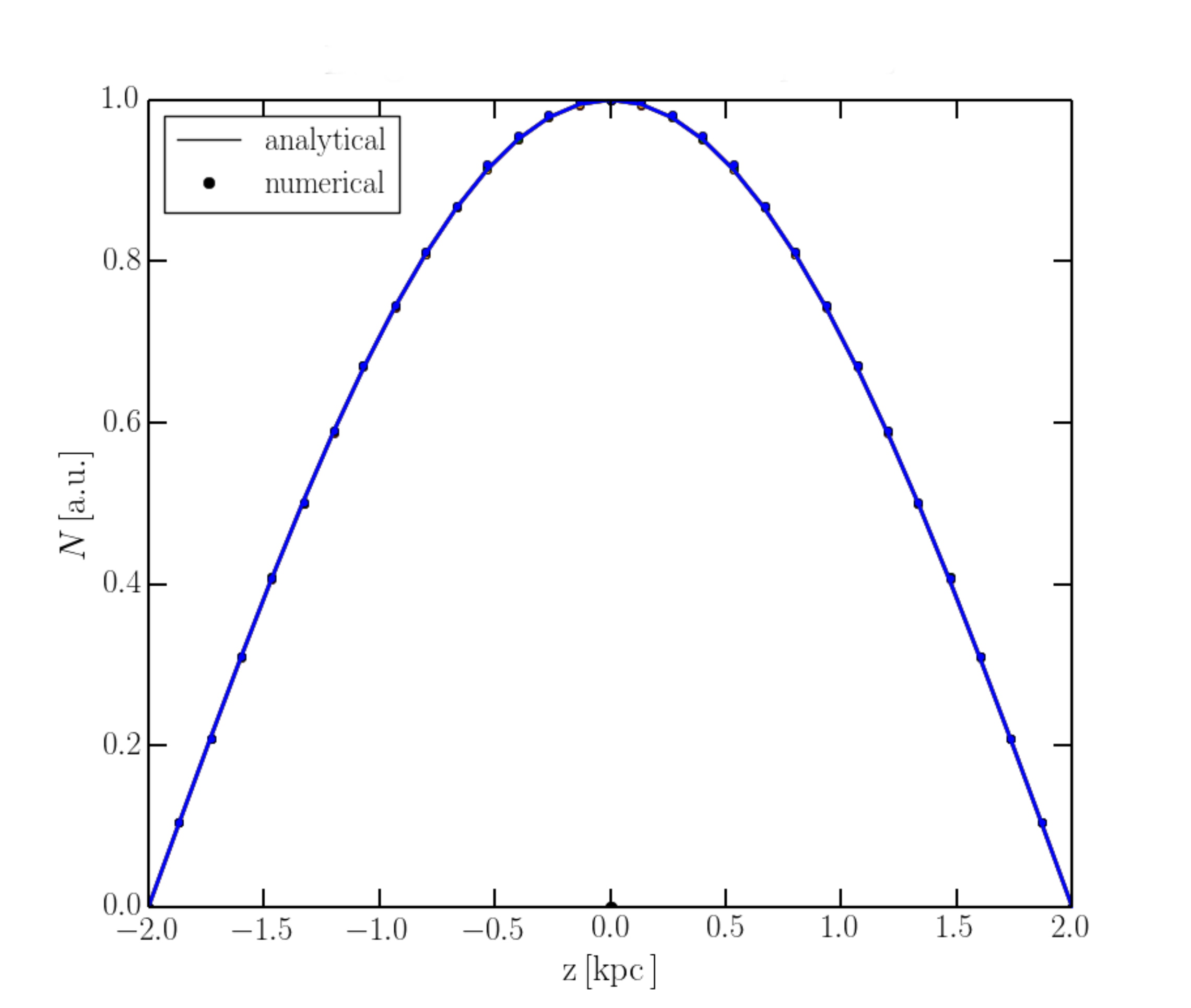}
  \includegraphics[width=0.45\textwidth]{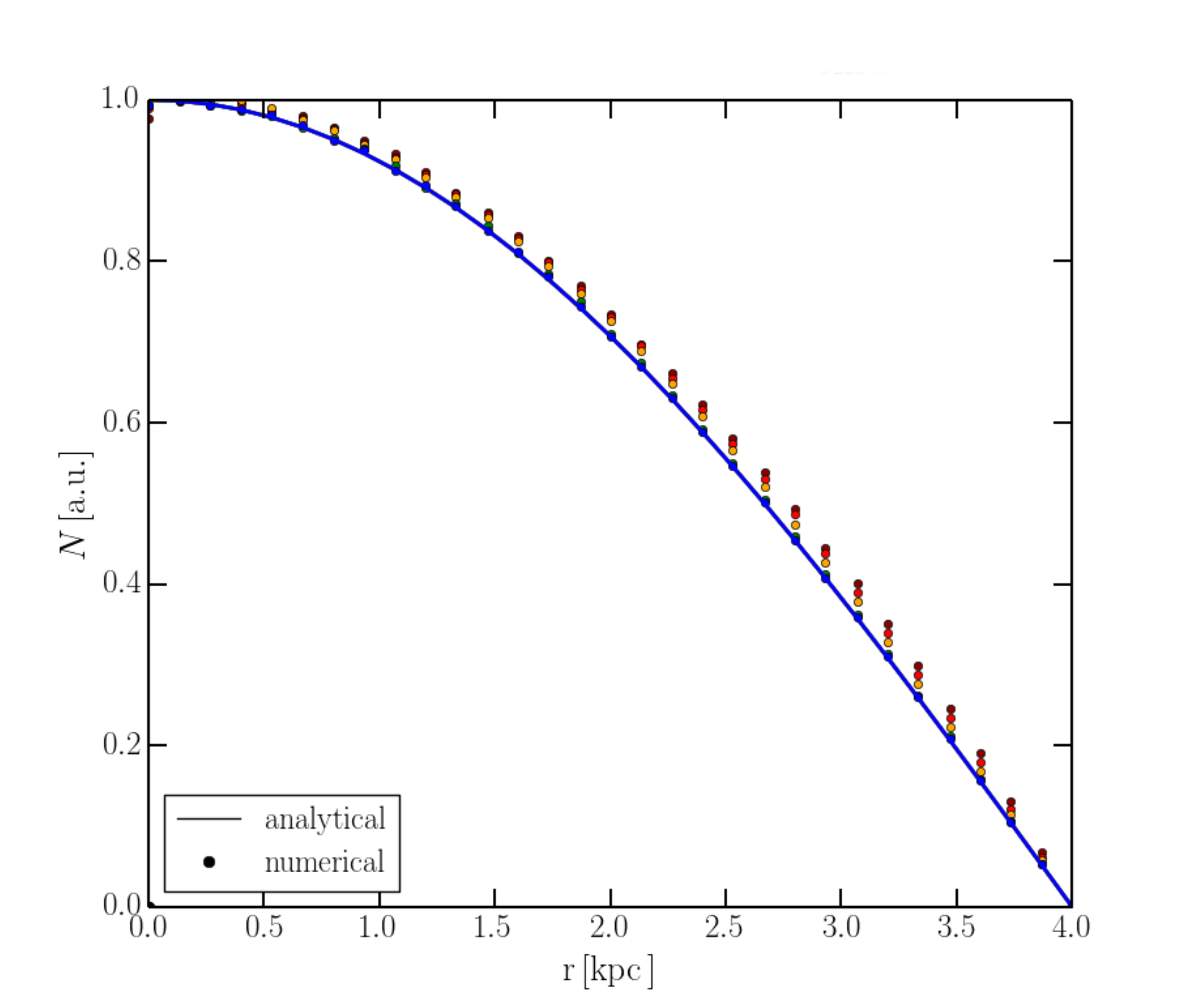}
 \caption{Test P1: comparison between numerical (circles) and analytical (solid lines) solution along $z$ (left) and along $R$ (right) at different times (colors): $10$ Myr (dark red), $20$ Myr (red), $40$ Myr (yellow), $100$ Myr (green) and $1$ Gyr (blue).}
 \label{fig:testP1}
\end{figure}

\begin{figure}[h]
  \center
  \includegraphics[width=0.45\textwidth]{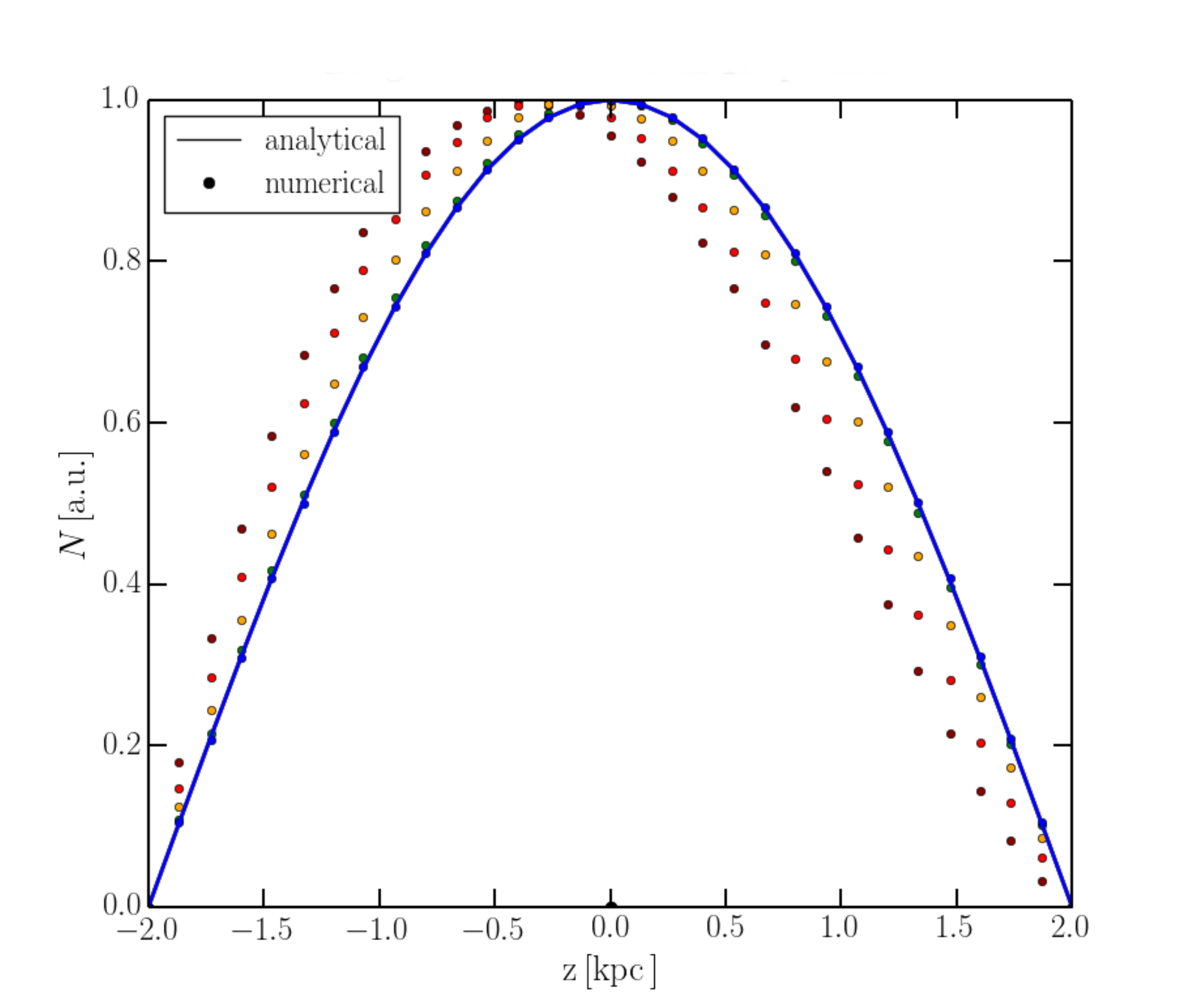}
  \includegraphics[width=0.45\textwidth]{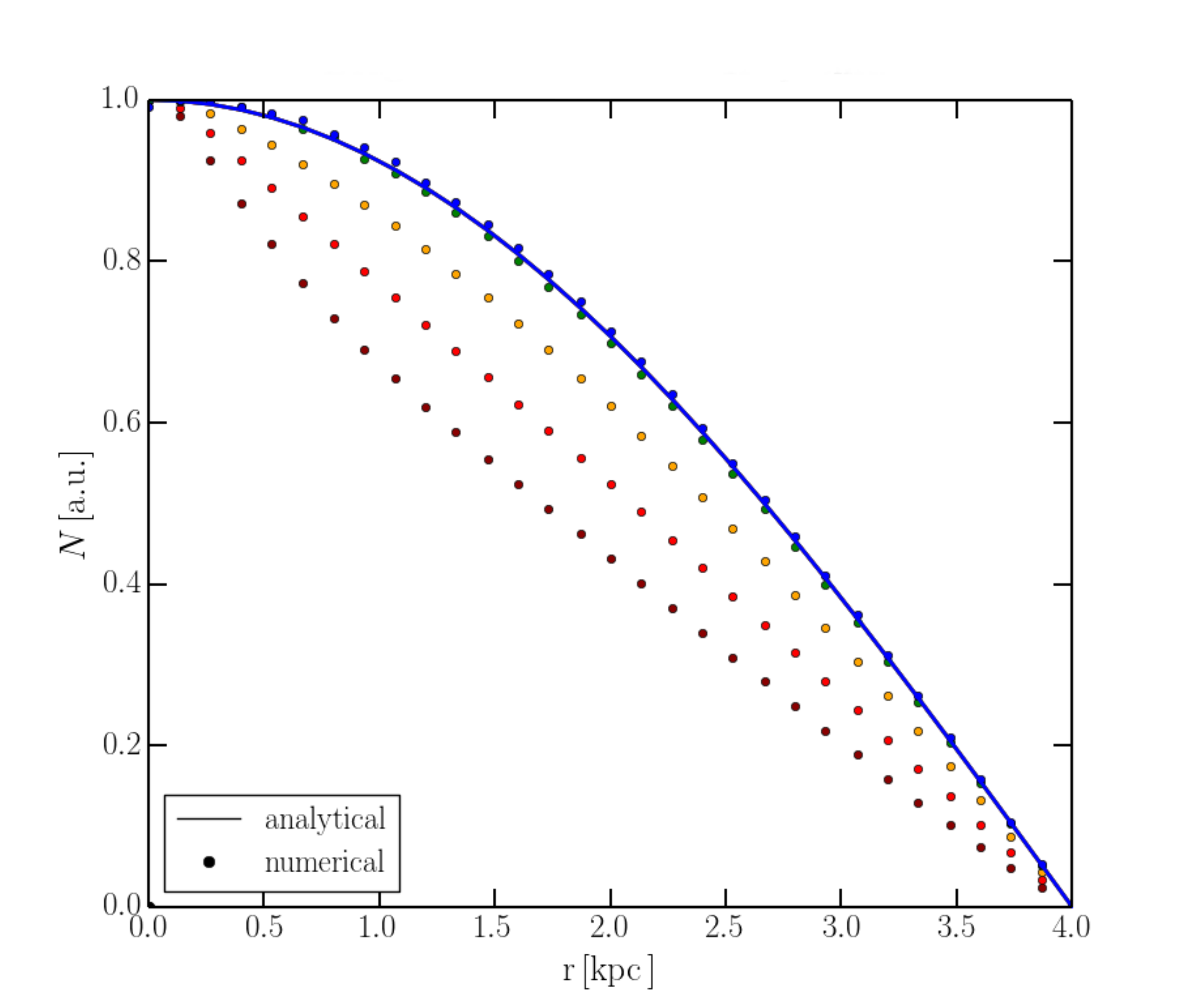}
 \caption{Test P2: comparison between numerical (circles) and analytical (solid lines) solution along $z$ (left) and along $R$ (right) at different times (colors): $10$ Myr (dark red), $20$ Myr (red), $40$ Myr (yellow), $100$ Myr (green) and $1$ Gyr (blue).}
 \label{fig:testP2}
\end{figure}

\begin{figure}[!h]
  \center
  \includegraphics[width=0.45\textwidth]{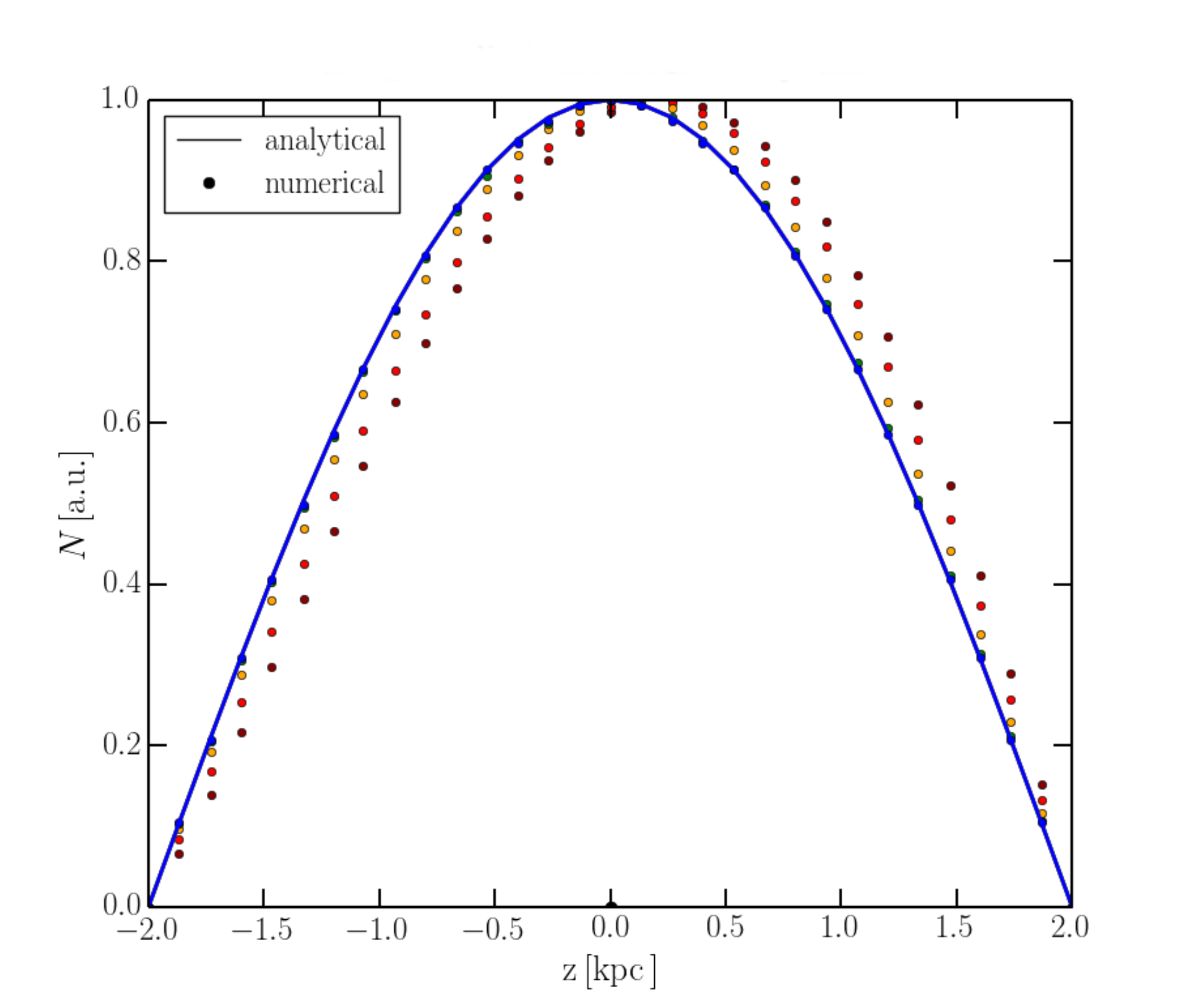}
  \includegraphics[width=0.45\textwidth]{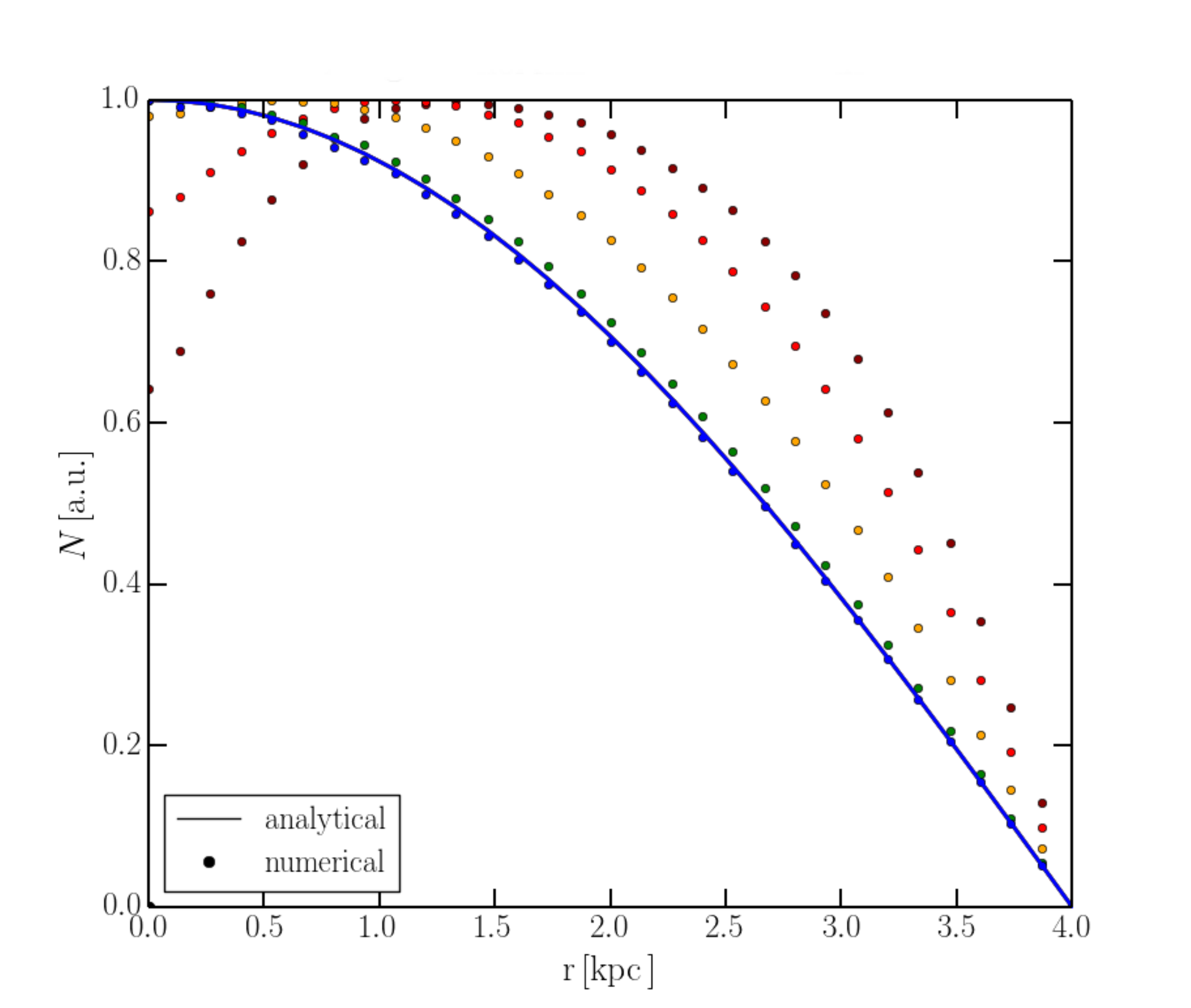}
 \caption{Test P3: comparison between numerical (circles) and analytical (solid lines) solution along $z$ (left) and along $R$ (right) at different times (colors): $10$ Myr (dark red), $20$ Myr (red), $40$ Myr (yellow), $100$ Myr (green) and $1$ Gyr (blue).}
 \label{fig:testP3}
\end{figure}

In Table~\ref{tab:tab2}, the parameters used for the tests are reported. The results of these tests are shown in Figs.~(\ref{fig:testP1})--(\ref{fig:testP3}). The numerical solution $N$ converges quite rapidly to the analytical $\psi$ given in eq.~(\ref{eq:PicardPsi}). Again, the agreement is remarkable.


\begin{table}[t]
 \center
 \begin{tabular}{ccc|ccccccccccccc|cccccccc}
  \toprule
    & Run & & & ($N_R$,$N_z$) & & $N_p$ & & $dt$ [Myr] & & $B_R$ & & $B_z$ & & $D_{0\perp}/D_{0\|}$\\
   \hline 
    & P1 & & & (31,31) & & 2 & & 0.1 & & 0 & & $B_0$ & & 0.05\\
    & P2 & & & (31,31) & & 2 & & 0.1 & & $B_0$ & & $B_0$ & & 0.05\\
    & P3 & & & (31,31) & & 2 & & 0.1 & & $B_0/2$ & & $B_\infty+B_0\exp(-r^2/L_B^2)$ & & 0.05\\
   \hline
  \hline
 \end{tabular}
   \caption{Simulation parameters for the comparison with analytical solutions. The constants in physical units are $D_{0\|}=10^{28}$ cm$^2$/s, $B_0=10$ $\mu$G, $B_\infty=1$ nG, and $L_B=2$ kpc.
   }
   \label{tab:tab2}
\end{table}


\clearpage

\providecommand{\href}[2]{#2}\begingroup\raggedright\endgroup

\end{document}